\documentclass[preprint2]{aastex}
\usepackage{url}

\begin{document}

\title{\emph{Swift} X-ray Telescope Monitoring of \emph{Fermi}-LAT Gamma Ray Sources of Interest}
\author{Michael C. Stroh \& Abe D. Falcone}
\affil{Department of Astronomy \& Astrophysics, \\The Pennsylvania State University, University Park, PA 16802}

\begin{abstract}
We describe a long-term \emph{Swift} monitoring program of \emph{Fermi} gamma-ray sources, particularly the 23 gamma-ray "sources of interest."  We present a systematic analysis of the \emph{Swift} X-ray Telescope light curves and hardness ratios of these sources, and we calculate excess variability.  We present data for the time interval of 2004 December 22 through 2012 August 31. We describe the analysis methods used to produce these data products, and we discuss the availability of these data in an online repository, which continues to grow from more data on these sources and from a growing list of additional sources.  This database should be of use to the broad astronomical community for long term studies of the variability of these objects and for inclusion in multi-wavelength studies. \\
\end{abstract}

\section{Introduction} 
\label{sec:introduction}
Prior to the launch of the \emph{Fermi} Gamma Ray Space Telescope, a list of 23 "sources of interest" was created as the initial list of sources to receive automated analysis, rapid release of public \emph{Fermi} data, and monitoring information.  These sources were deemed to be among the most likely candidates for interesting \emph{Fermi} Large Area Telescope (LAT) \citep{atwood2009} high states and/or intense studies with multiwavelength campaigns.  These sources are all active galactic nuclei, with the exception of one X-ray/TeV binary, LS I +61 303.  As the mission progressed, this list grew into a more extensive monitored source list, which leads to monitoring data on any additional sources that exhibit a high state in \emph{Fermi}-LAT corresponding to $>1\times10^{-6}$ ph cm$^{-2}$ s$^{-1}$ in the $100$ MeV $<$ E $<300$ GeV band.

The \emph{Fermi}-LAT monitored sources, particularly the "sources of interest" that are dominated by active galactic nuclei, include some of the most interesting sources in high energy astronomy.  They are often observed through multi-wavelength campaigns that can span from radio to TeV gamma-rays.  These sources typically exhibit a spectral energy distribution that contains at least two distinct components, with a large bump due to electron synchrotron emission peaking in the optical to X-ray bands and another large bump that typically peaks in the gamma ray band \citep{abdo2010}.  This second bump is typically modeled with a dominant contribution from leptonic processes, namely inverse Compton emission with seed photons from either synchrotron emission or from surrounding regions, but it is sometimes modeled with strong contributions from hadronic processes, namely proton synchrotron emission and/or proton cascades.  The study of these different emission processes is important for furthering our understanding of the particle acceleration in these dynamic environments.  These "sources of interest" provide some of the best opportunities to perform these studies and to determine the ratios of hadron and lepton acceleration.  Broadband monitoring at many wavelengths, particularly the X-ray band where the synchrotron peak often lies and the gamma-ray band where the high energy peak lies, is required to further these studies.  

\emph{Swift} \citep{gehrels04} has a flexible observing strategy and the ability to react quickly to target of opportunity requests (ToOs), making it a great asset in multiwavelength campaigns. \emph{Swift} is equipped with a co-pointed X-ray Telescope (XRT), a Ultraviolet/Optical telescope (UVOT), and a hard X-ray telescope (Burst Alert Telescope; BAT), providing inherent multiwavelength coverage.  In this paper, we report on monitoring observations with \emph{Swift}-XRT \citep{burrows05} and on the characteristics of extensive real-time light curve data that are available to all observers.  

\section{Data and Analysis}
\label{sec:dataAnalysis}
The data set described in this paper is comprised of 5.4$\times10^{6}$ seconds of observing time from over 3100 observations on the original \emph{Fermi}-LAT 23 sources of interest. These data were obtained between 2004 December 22 to 2012 August 31.  Much of these data were obtained through weekly 1ks observations over four month periods when the sources could be observed by ground-based observers, and some of these data were obtained through other multiwavelength campaigns, particularly on heavily observed sources such as Mrk 421 and Mrk 501. Since there are also times when an unrelated observation leads to serendipitous observations of one of these sources, our data products include every observation performed within $5\arcmin$ of the source positions. Since the XRT has to rely on passive cooling in order to keep the temperature of the detector below -50C, the observatory sometimes points at locations in the sky that will minimize the Earth elevation angle and decrease the temperature \citep{kennea05}. When these pointings include the monitored sources, we include them in our data set using the source position selection criteria mentioned above. Our data contain observations from Target of Opportunity requests, Guest Investigator programs, calibrations, and observations taken from BAT triggers on or near the sources.

The \emph{Swift}-XRT data were processed using the most recent versions of the standard \emph{Swift} tools: \emph{Swift} Software version 3.9 and FTOOLS version 6.12 \citep{blackburn95}. Observations are processed separately using xrtpipeline version 0.12.6. Hot and flickering pixels are initially removed using xrthotpix and hot pixels are additionally removed by rejecting data where the XRT's CCD temperature is $\ge-47C$. We use the standard grade selections of 0-12 and 0-2 in the Photon Counting (PC) and Windowed Timing (WT) modes respectively. Since the spacecraft attitude information is less accurate while in WT settling mode, only pointing mode data are used.

\subsection{Light Curves and Hardness Ratios}
\label{sec:lightCurves}
Light curves are generated using xrtgrblc version 1.6. In PC mode, circles are used to describe the source regions except when the source count rate is $>0.5\enskip counts\enskip s^{-1}$, in which case annuli are used. Annuli are used to describe the background regions for all PC mode data. The radii of the PC regions depend on the current count rate and the sizes are summarized in Table \ref{pcregions}. Rectangles of $20\arcsec$ height are used to describe source, pile-up and background regions for all WT mode data. The widths of the source, pileup, and background regions depend on the count rate and are summarized in Table \ref{wtregions}. Since background regions are chosen indiscriminately by xrtgrblc regardless of the field, there may be rare cases where additional sources may reside in the background regions but our online repository described in Section \ref{sec:onlineRepository} provides summed PC images so that the user can be aware of these issues in the unlikely event that they arise.

In order to handle both piled-up observations and cases where the sources land on bad columns, vignetting and point spread function correction is handled using xrtlccorr. Slight pile-up has been observed in cases where the count rate is $>100~cts~s^{-1}$ \citep{romano06}, but xrtgrblc doesn't use a pile-up region until $300~cts~s^{-1}$; however, Mrk 421, Cir X-1, Cyg X-3, GX 304-1 and 1A 0535+262 are the only sources in our sample where this small systematic effect may occur since they are the only sources with any observations in the $>100~cts~s^{-1}$ count rate regime. In these cases, the effect on the flux light curve is unmeasurably small compared to the statistical error bars. Additionally, at $400~cts~s^{-1}$ the background and source regions are no longer disjoint which could lead to some slight source contamination in the background regions, but 1A 0535+262 is the only source in our sample with a count rate that high. More information on xrtgrblc can be found in its help file at \url{http://heasarc.gsfc.nasa.gov/lheasoft}\\ \url{/ftools/headas/xrtgrblc.html}.

The overall light curves use a bin size of one observation per bin and contain all \emph{Swift}-XRT observations in PC and WT modes from 2004 December 22 through 2012 August 31. Individual observations have typical durations of 1-4 ksec.  Single observation light curves provide more detail by using a step binning procedure so that if the count rate is less than 0.4, 1, 10, 100 and 10000 $counts\enskip s^{-1}$, then 20, 50, 200, 1000 and 2000 $counts\enskip bin^{-1}$ are used respectively. In cases where the single observation light curve produces an upper limit but the overall light curve  produces a non-upper limit, the single observation light curve will use the overall light curve bin. Finely binned overall light curves contain all bins from the single observation light curves.

The hardness ratio is defined as $R_2$/$R_1$, where $R_2$ is the rate in the 2-10 keV band and $R_1$ is the rate in the 0.3-2 keV band.  Each band is binned with the same bin size, using the same methods that are used for the light curves; the long term $R_2$/$R_1$ curves receiving a single bin per observation, and the detailed curves receive the stepped binning described above with the hard band rate, $R_2$, defining the rate that is used to define the number of counts per bin.  Three sigma upper limits are used for light curves and hardness ratios, and all error bars are reported at the 1-sigma level.

Bright Earth limb or an increased number of hot pixels can occasionally cause the XRT to switch between PC and WT modes. Mode switching in the light curves is filtered out by rejecting WT bins with less than 15 raw counts in the source region. WT hardness ratio bins are rejected if they do not temporally overlap at least one WT bin in the light curve.

\subsection{Spectral Fitting}
\label{sec:spectralFitting}
In order to calculate count rate to flux conversion factors, spectral fits were generated using XSPEC version 12.7.1 \citep{arnaud96}. Every observation in the overall light curves that resulted in a source detection was broken up into individual snapshots. Source and background information was extracted from each snapshot using the same region definitions used by xrtgrblc. Ancillary response files (ARFs) were created using xrtmkarf version 0.6.0, and exposure maps were used for PSF corrections. ARFs were then combined using addarf 1.2.6 and were weighed by the number of counts in the source regions. 

Source and background regions were combined as well as ARFs over the full time domain for which the response matrix files (RMFs) were applicable, and data for each mode were fit independently. This resulted in subsets of data that were broken up into time periods during which different ARFs and RMFs applied.  Data were binned using grppha version 3.0.1 with a minimum of 20 counts per bin.

Since most of the WT data taken for PKS 0235+164, LS I +61 303, S5 0716+714, W Com, 3C 279, 1Jy 1406-076, PKS 1510-089, BL Lacertae, and 1ES 2344+514 were due to mode switching, we didn't consider any WT data for those sources.

In our goal to be conservative with our spectral fitting, we removed the relatively bright sources 2XMM J140854.1-075323 and 1WGA J1221.6+2811 from the background regions of 1Jy 1406-076 and W Com respectively. Using the sosta routine in XIMAGE we found that of these sources have count rates of around $4 \times 10^{-3}~counts~s^{-1}$ which is about an order of magnitude less than the typical count rates observed by 1Jy 1406-076 and W Com. We used 12\arcmin circular regions to remove these two background sources which seemed to be large enough to contain their PSFs while not encroaching upon the sources of interest. 

Some remaining flickering pixels that weren't successfully filtered out by xrtpipeline caused an excess in the edges of the spectrum, so we ignored these regions of the spectra. In some of our longer data sets, the background is dominated by the Ni $K_{\alpha}$ line at $7.478~keV$ due to flourescence from the telescope material \citep{moretti09}. To be safe we ignored the 7.0 - 10.0 keV band of the spectrum to account for the Ni $K_{\alpha}$ line and the potential for flickering pixels.

The spectra were fit using $\chi^2$ statistics assuming an absorbed power law with freely varying $N_H$, power law photon index, and normalization constant. Since some of the spectral fits for W Com, 3C 273, 1Jy 1633+38, Mrk 501, and PKS 1730-130 resulted in low $N_H$ values, we set the galactic $N_H$ \citep{kalberla05} to be a lower bound for these spectral fits. For the full set of fits, the average $\chi^2/d.o.f. = 1.08 \pm 0.03$.  Table \ref{specFitParameters} displays the averaged best fit parameters for each source. After performing a spectral fit for an absorbed power-law on the subset of the data, we calculated the unabsorbed flux for the spectrum along the full 0.3 - 10.0 keV energy range to match the energy range used to create the light curves. The time period over which the unabsorbed flux was calculated was also used in calculating the weighted mean count rate. These together were used to create individual count rate to flux conversion factors.  The mean of the individual conversion factors was used to calculate a final count rate to flux conversion factor for each source, which could then be applied to the rate light curves to create flux light curves. In cases for which the spectral shape variability becomes large, these average flux-to-rate conversion factors given in table \ref{sourcesFlux} should be viewed as approximations.

\subsubsection{Mrk 501 and 3C 273}
In the case of the brightest sources that are monitored frequently, the RMFs couldn't correctly compensate for the O-edge at 0.54 keV during WT mode observations. For these observations, the O-edge was removed from the fits of the spectra by ignoring the $<$0.8 keV energy bins.

\subsubsection{Mrk 421}
Mrk 421 is the brightest and most frequently observed source in our list which caused similar issues to those of Mrk 501 and 3C 273. Furthermore, the brightness and variability of Mrk 421 motivates a spectral analysis with some time resolution.  Instead of removing more edges from this spectra, we split the time domain and calculated spectral parameters and flux conversion factors in these separate time regions.

\section{Results}
\label{sec:results}
We created rate light curves for the period from December 2004 through August 2012. Light curves in a rate format can be found in our online repository described in Section \ref{sec:onlineRepository}. We used the rate light curves and the spectral fits described in Section \ref{sec:spectralFitting} to create flux light curves in Section \ref{sec:fluxLightCurves}, and the rate light curves were used to conduct an excess variability study in Section \ref{sec:excessVariance}.
 
\subsection{Excess Variance}
\label{sec:excessVariance}

In order to probe the variability of these sources, we calculated the excess variance \citep{nandra97} for each source as defined by 

\begin{equation} \label{f1}
\sigma_{rms}^{2} \equiv \frac{1}{N\mu^2} \sum_{i=1}^N [(X_i - \mu)^2-\sigma_i^2]
\end{equation}

where $\mu$ is the unweighted average of the count rates, $X_i$, with errors, $\mu_i$. The excess variabilities for the overall, coarsely-binned light curves ($\sigma_{rms1}^2$) and for the finely binned light curves ($\sigma_{rms2}^2$) can be found in Table \ref{sources}.

PKS 0235+164 and W Com have the highest $\sigma_{rms1}^2$ and $\sigma_{rms2}^2$. Each of these sources has a moderately low mean flux, and each has variability across nearly two orders of magnitude.  PKS 0235+164 has a long-term light curve with a remarkably isolated period of increased count rate.

1Jy 1406-076 has the lowest $\sigma_{rms1}$ and $\sigma_{rms2}$, and this source also has the lowest mean count rate, $(1.94 \pm 0.06) \times 10^{-2} \enskip counts \enskip s^{-1}$.  It exhibits no major outbursts during the time period of this monitoring. Since most observations are only 1ks in length, the bins in the overall light curve for 1Jy 1406-076 have relatively large error bars. In addition, there are only a few cases where these observations were long enough for the finely binned light curves to produce more than a single bin, so during most observations of 1Jy 1406-076, the overall light curve and finely binned light curves are identical.

QSO B0827+243, 3C 273, PKS 1730-130, PKS 2155-304, 3C 454.3 and 1ES 2344+514 are the only sources where $\sigma_{rms2}^2 > \sigma_{rms1}^2$. Overall, $\sigma_{rms2}^2/\sigma_{rms1}^2 = 0.92 \pm 0.08$ which suggests that these sources are just as variable on short timescales as they are on the longer timescales.

\subsection{Flux Light Curves}
\label{sec:fluxLightCurves}
\begin{figure*}
\includegraphics[angle=270,width=15cm]{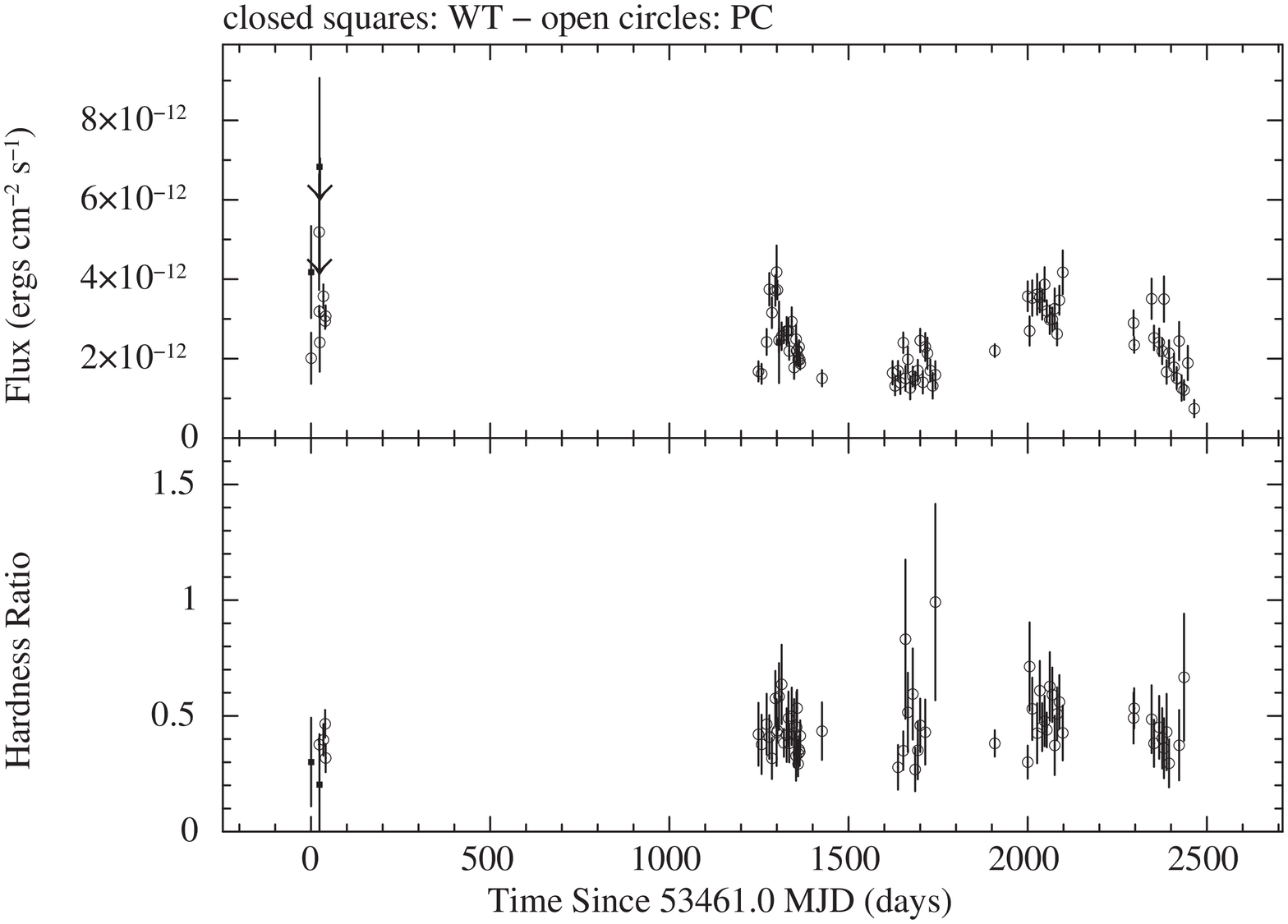}
\caption{Unabsorbed flux light curve and hardness ratio for PKS 0208-512. Three sigma upper limits are denoted by downward pointing arrows.}
\label{fig:pks_0208-512_light_curve}
\end{figure*}

\begin{figure*}
\includegraphics[angle=270,width=15cm]{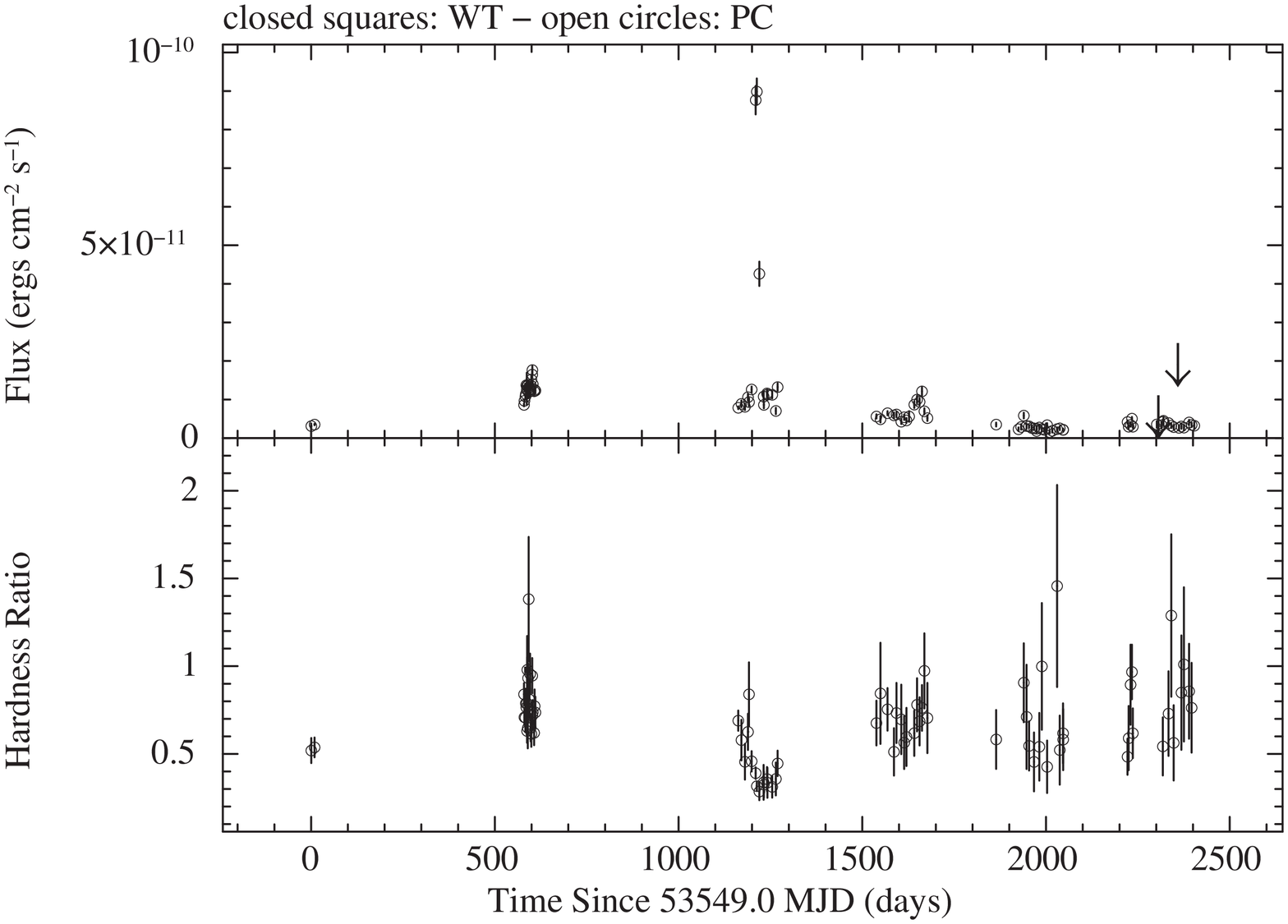}
\caption{Unabsorbed flux light curve and hardness ratio for PKS 0235+164. Three sigma upper limits are denoted by downward pointing arrows.}
\label{fig:pks_0235+164_light_curve}
\end{figure*}

\begin{figure*}
\includegraphics[angle=270,width=15cm]{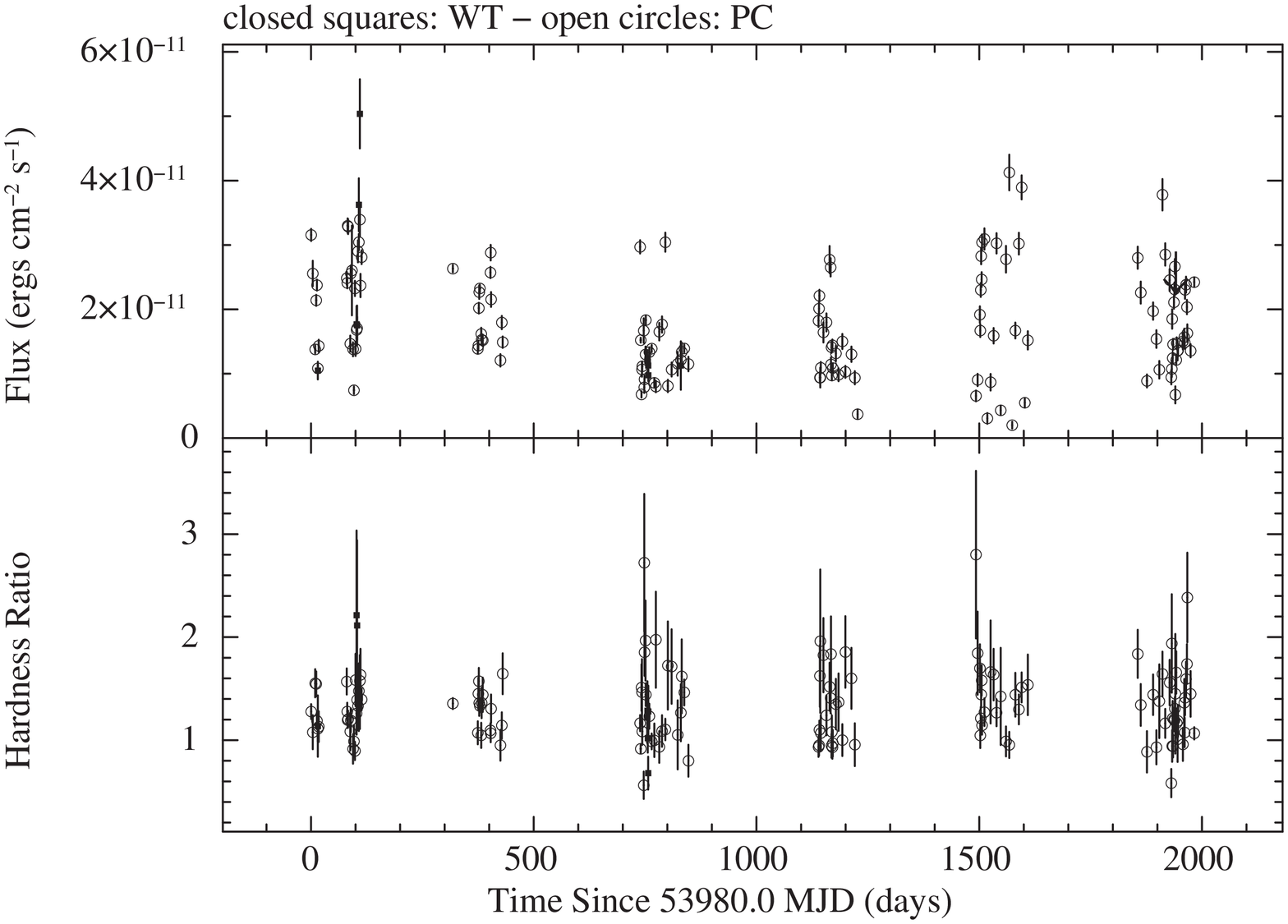}
\caption{Unabsorbed flux light curve and hardness ratio for LS I +61 303. Three sigma upper limits are denoted by downward pointing arrows.}
\label{fig:lsi+61_303_light_curve}
\end{figure*}

\begin{figure*}
\includegraphics[angle=270,width=15cm]{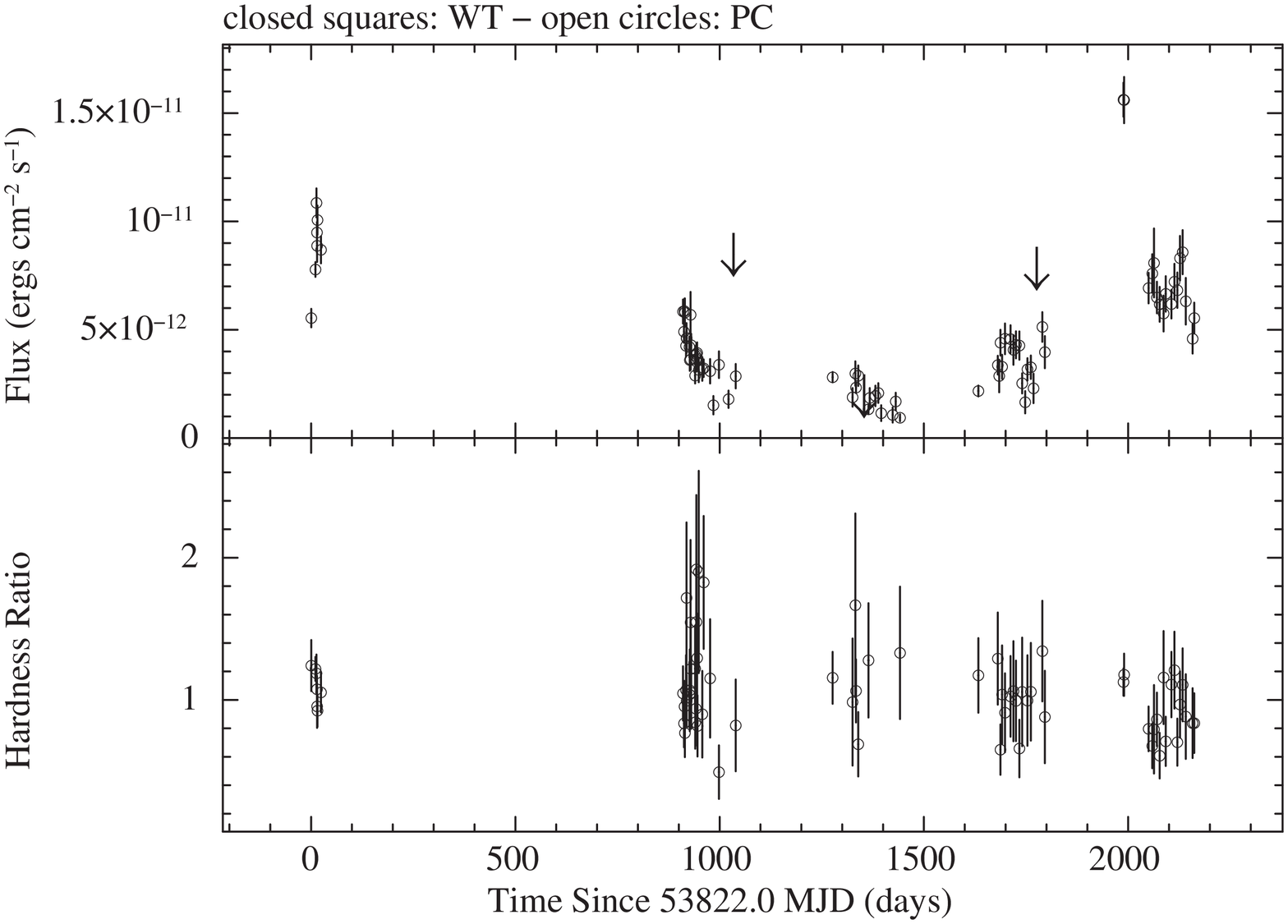}
\caption{Unabsorbed flux light curve and hardness ratio for PKS 0528+134. Three sigma upper limits are denoted by downward pointing arrows.}
\label{fig:pks_0528+134_light_curve}
\end{figure*}

\begin{figure*}
\includegraphics[angle=270,width=15cm]{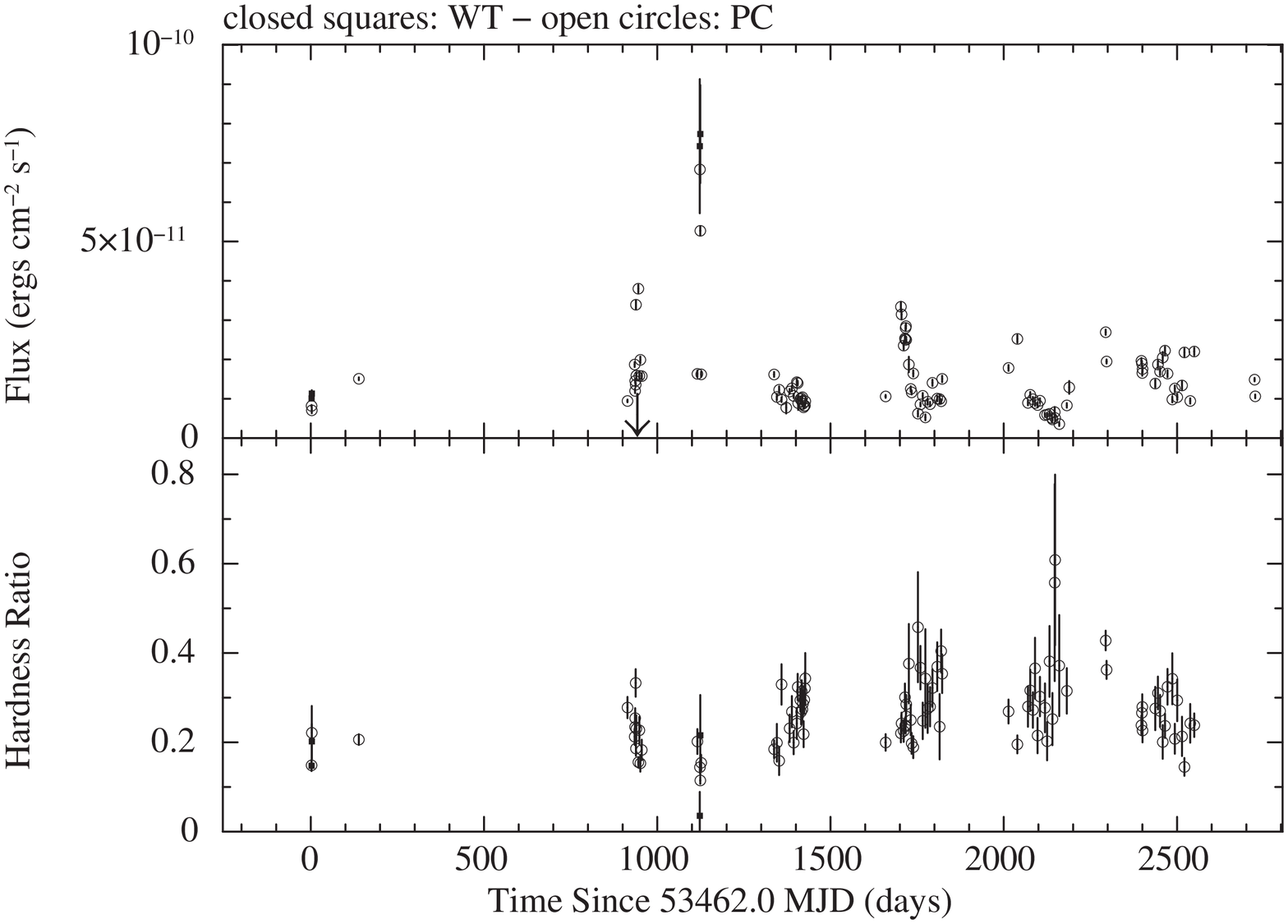}
\caption{Unabsorbed flux light curve and hardness ratio for S5 0716+714. Three sigma upper limits are denoted by downward pointing arrows.}
\label{fig:s5_0716+714_light_curve}
\end{figure*}

\begin{figure*}
\includegraphics[angle=270,width=15cm]{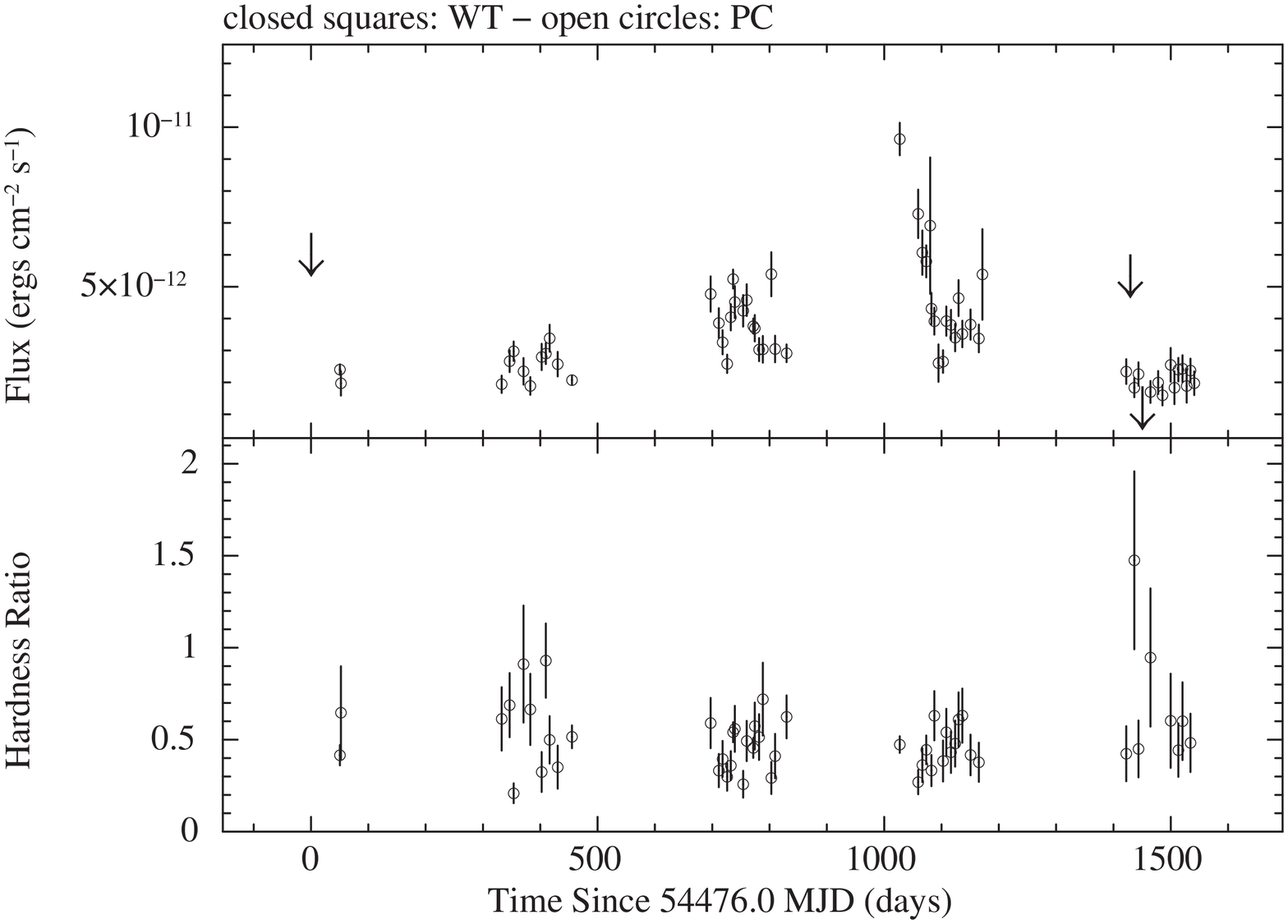}
\caption{Unabsorbed flux light curve and hardness ratio for QSO B0827+243. Three sigma upper limits are denoted by downward pointing arrows.}
\label{fig:qso_b0827+243_light_curve}
\end{figure*}

\begin{figure*}
\includegraphics[angle=270,width=15cm]{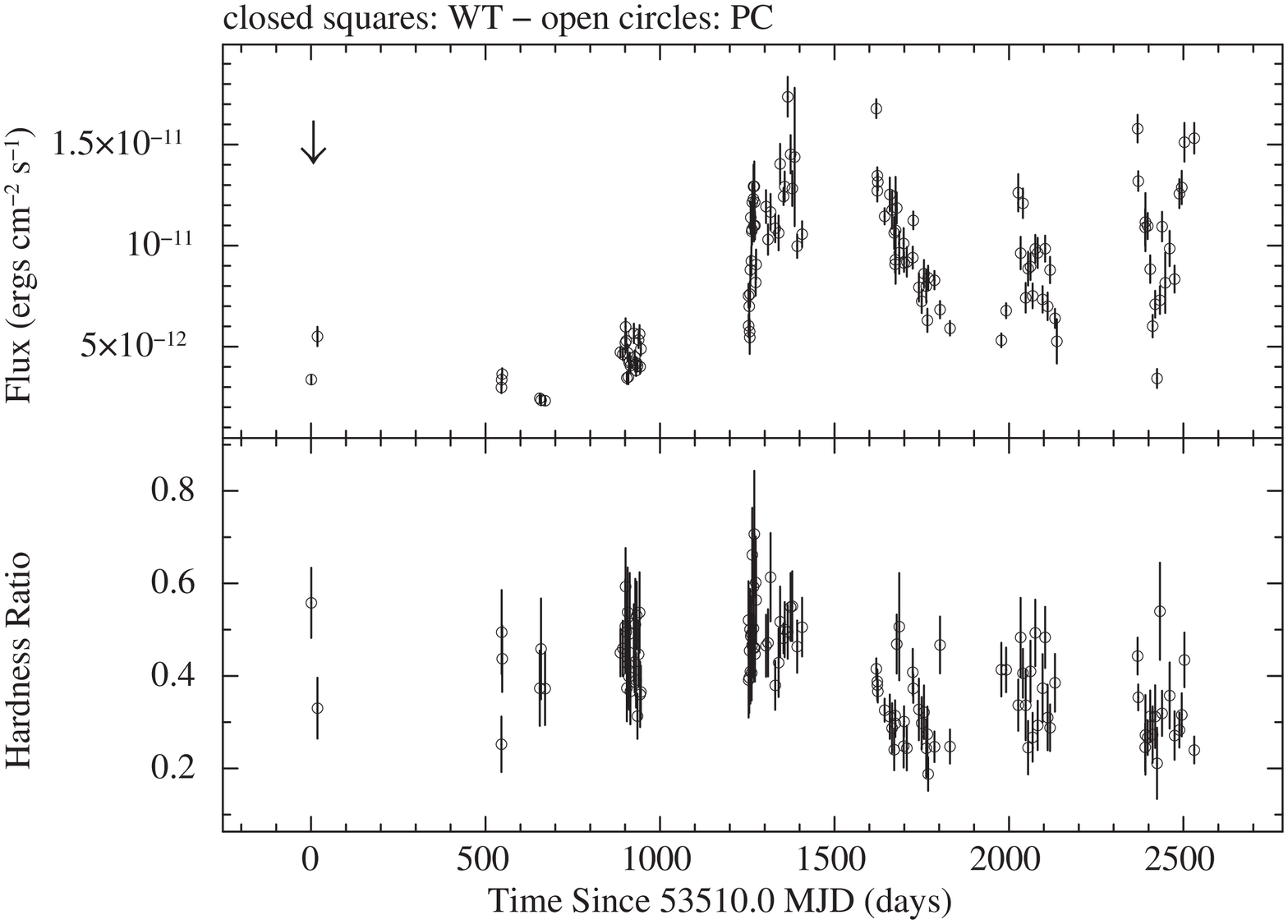}
\caption{Unabsorbed flux light curve and hardness ratio for OJ 287. Three sigma upper limits are denoted by downward pointing arrows.}
\label{fig:oj_287_light_curve}
\end{figure*}

\begin{figure*}
\includegraphics[angle=270,width=15cm]{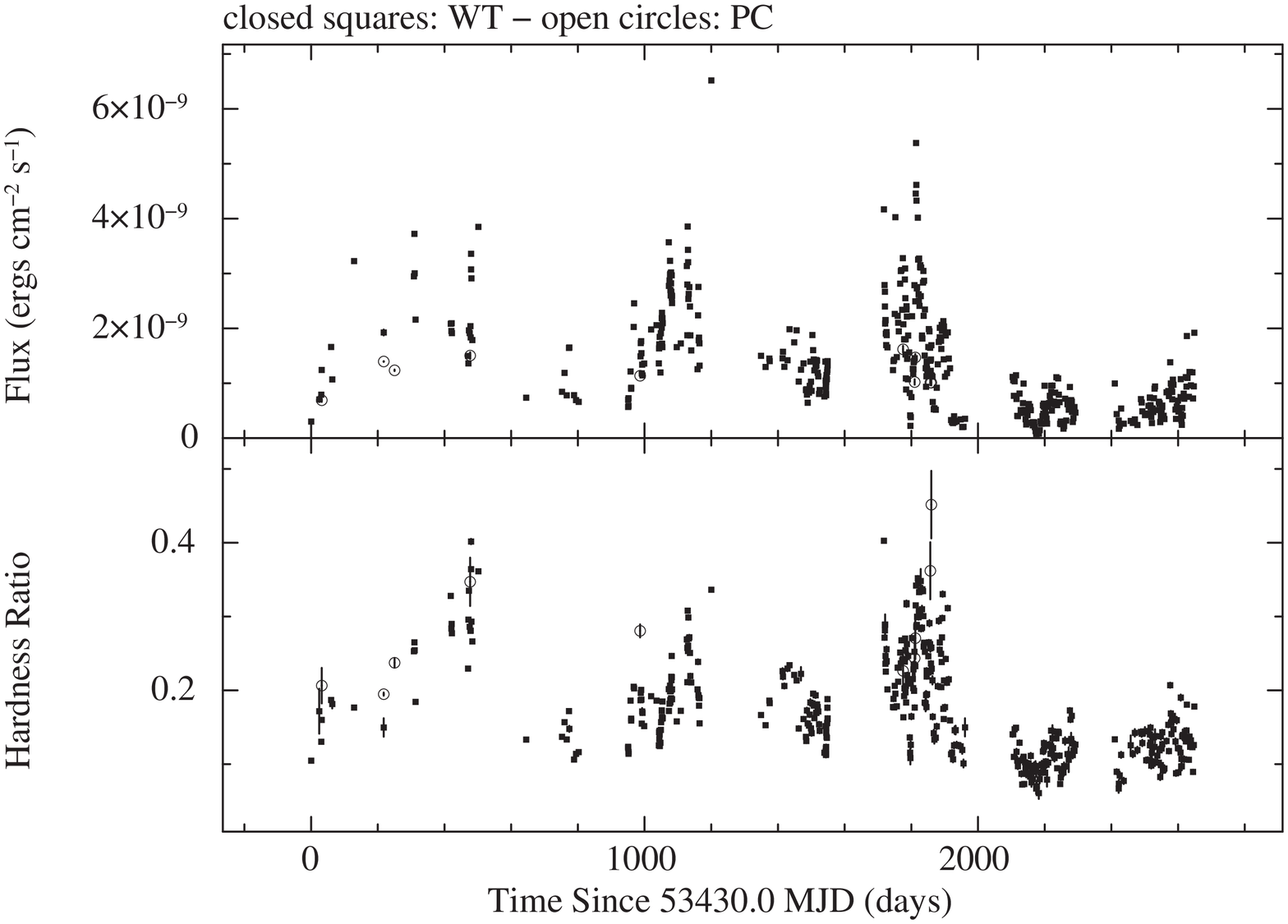}
\caption{Unabsorbed flux light curve and hardness ratio for Mrk 421.}
\label{fig:mrk_421_light_curve}
\end{figure*}

\begin{figure*}
\includegraphics[angle=270,width=15cm]{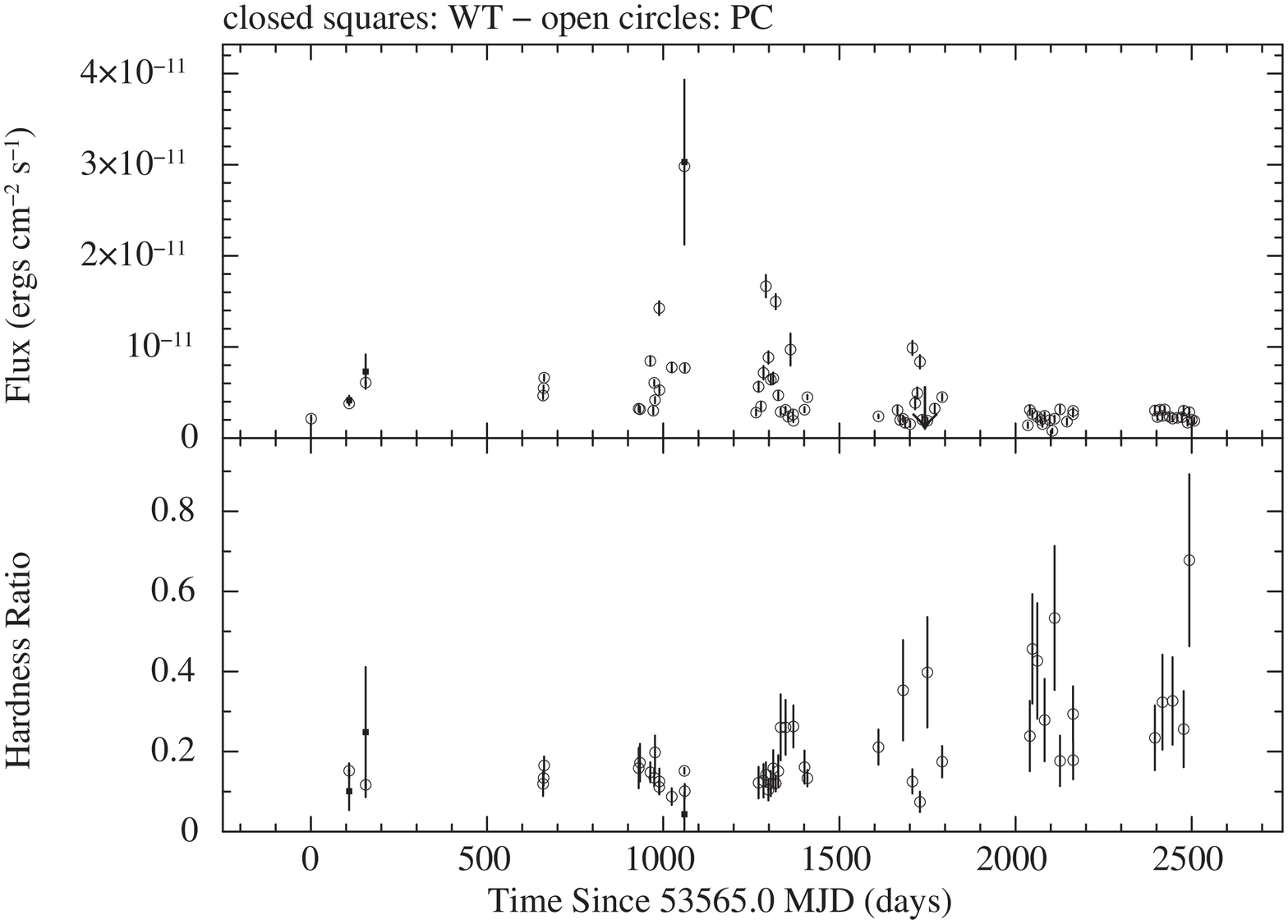}
\caption{Unabsorbed flux light curve and hardness ratio for W Com. Three sigma upper limits are denoted by downward pointing arrows.}
\label{fig:w_com_light_curve}
\end{figure*}

\begin{figure*}
\includegraphics[angle=270,width=15cm]{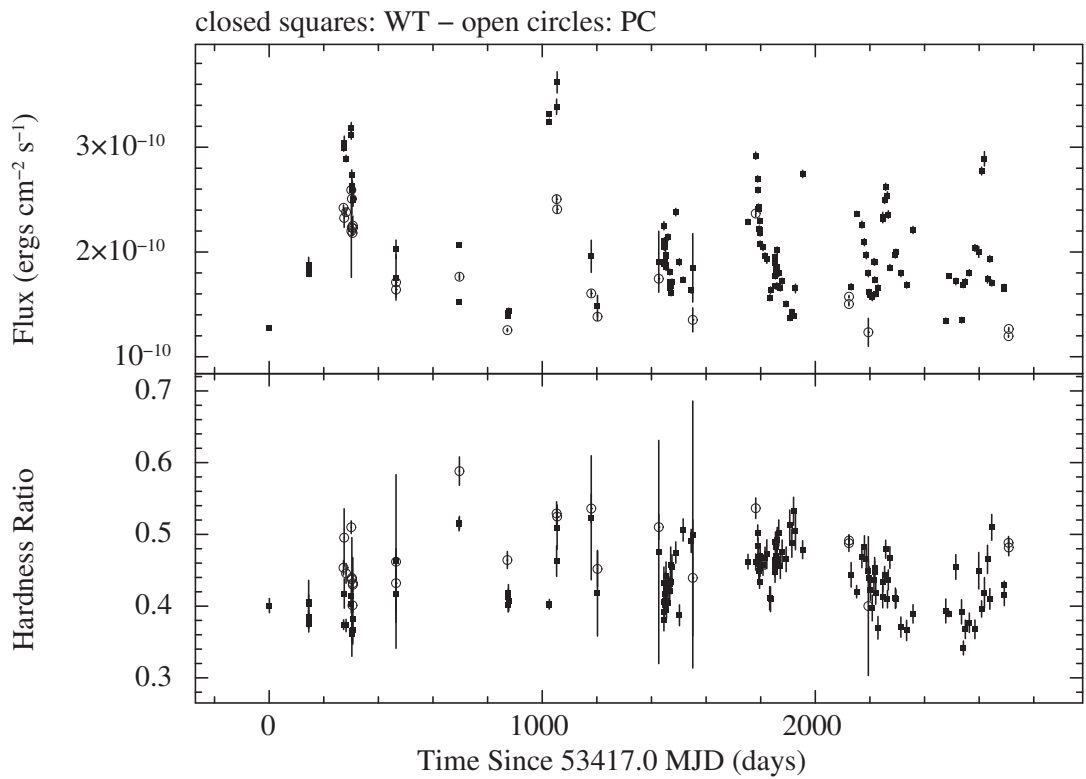}
\caption{Unabsorbed flux light curve and hardness ratio for 3C 273.}
\label{fig:3c_273_light_curve}
\end{figure*}

\begin{figure*}
\includegraphics[angle=270,width=15cm]{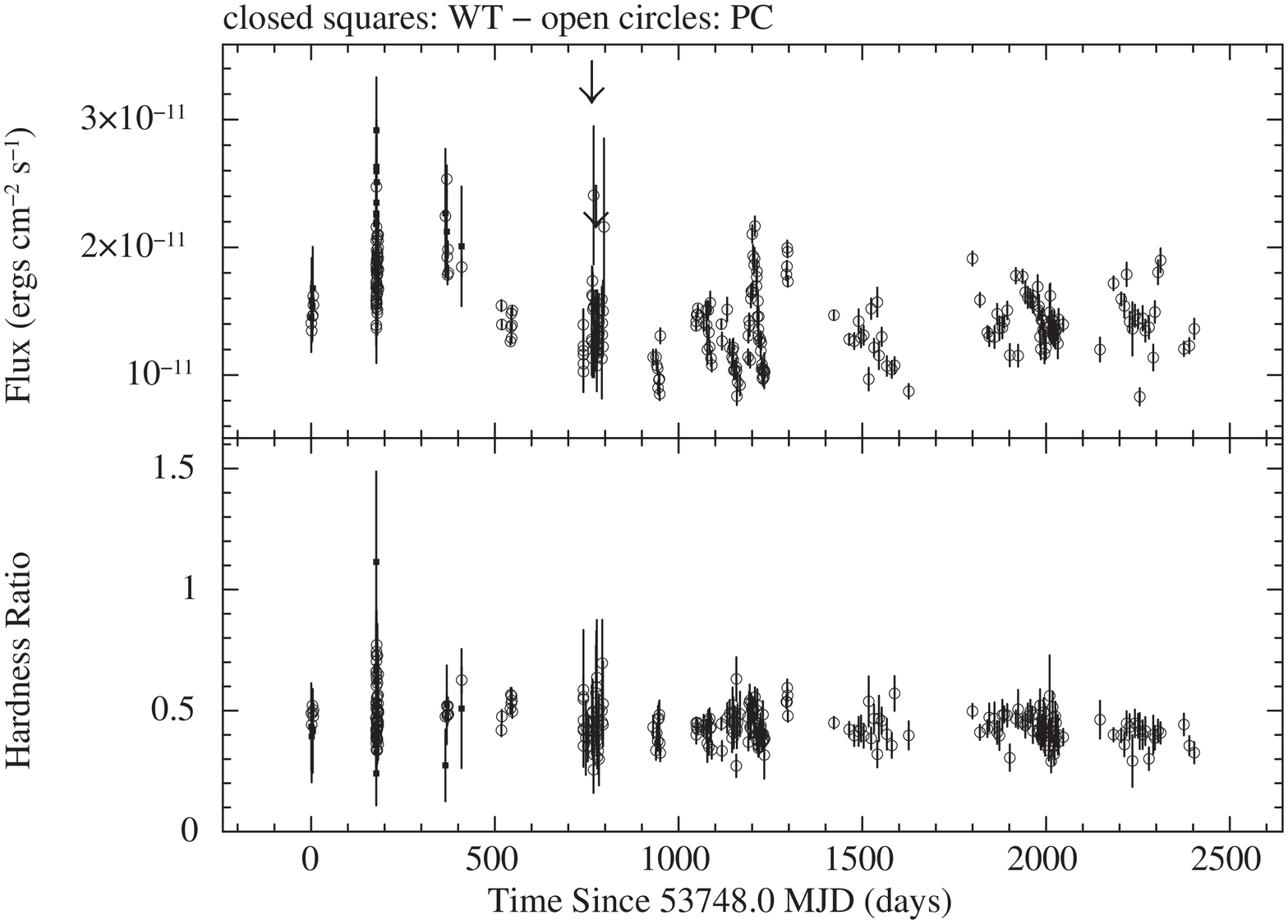}
\caption{Unabsorbed flux light curve and hardness ratio for 3C 279. Three sigma upper limits are denoted by downward pointing arrows.}
\label{fig:3c_279_light_curve}
\end{figure*}

\begin{figure*}
\includegraphics[angle=270,width=15cm]{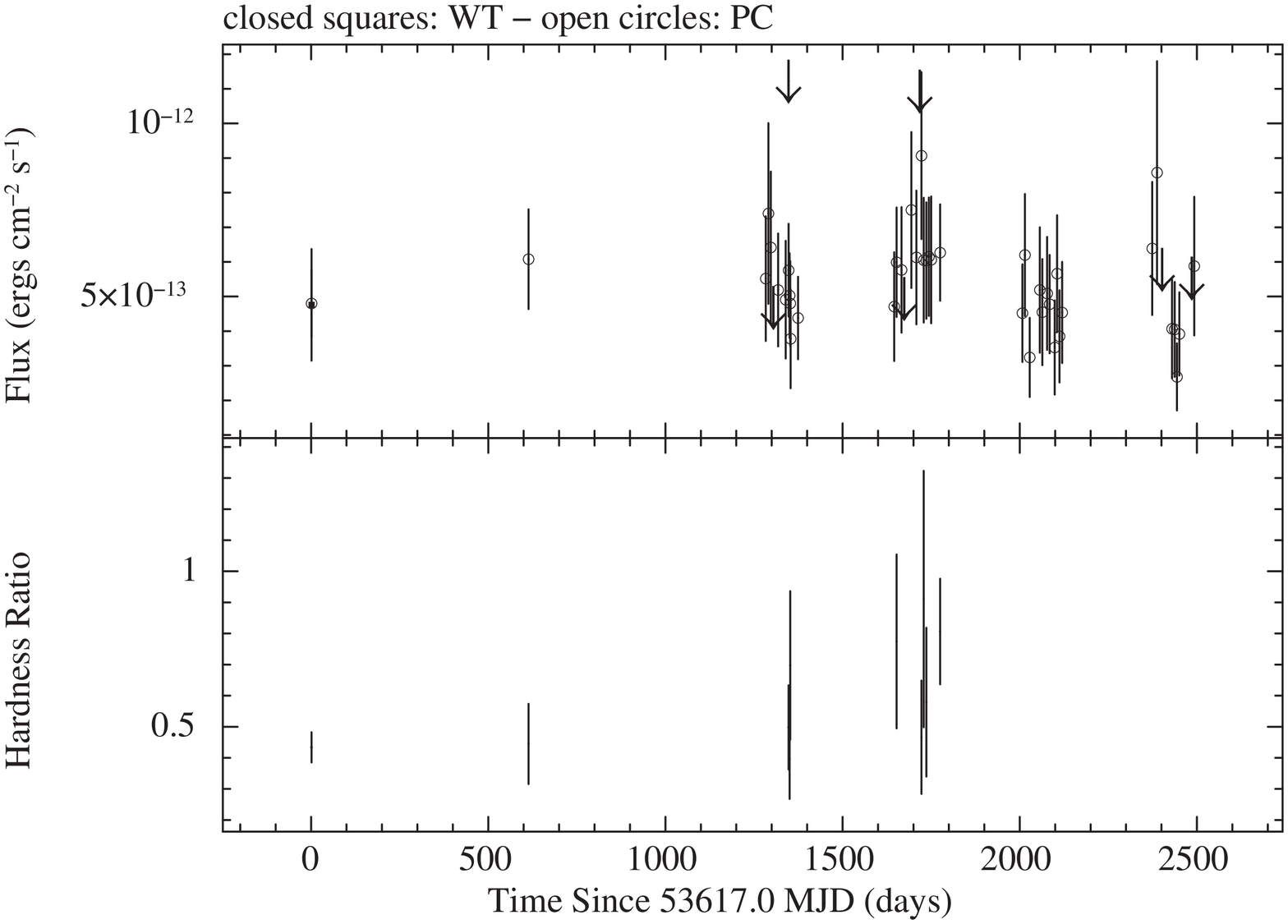}
\caption{Unabsorbed flux light curve and hardness ratio for 1Jy 1406-076. Three sigma upper limits are denoted by downward pointing arrows.}
\label{fig:1jy_1406-076_light_curve}
\end{figure*}

\begin{figure*}
\includegraphics[angle=270,width=15cm]{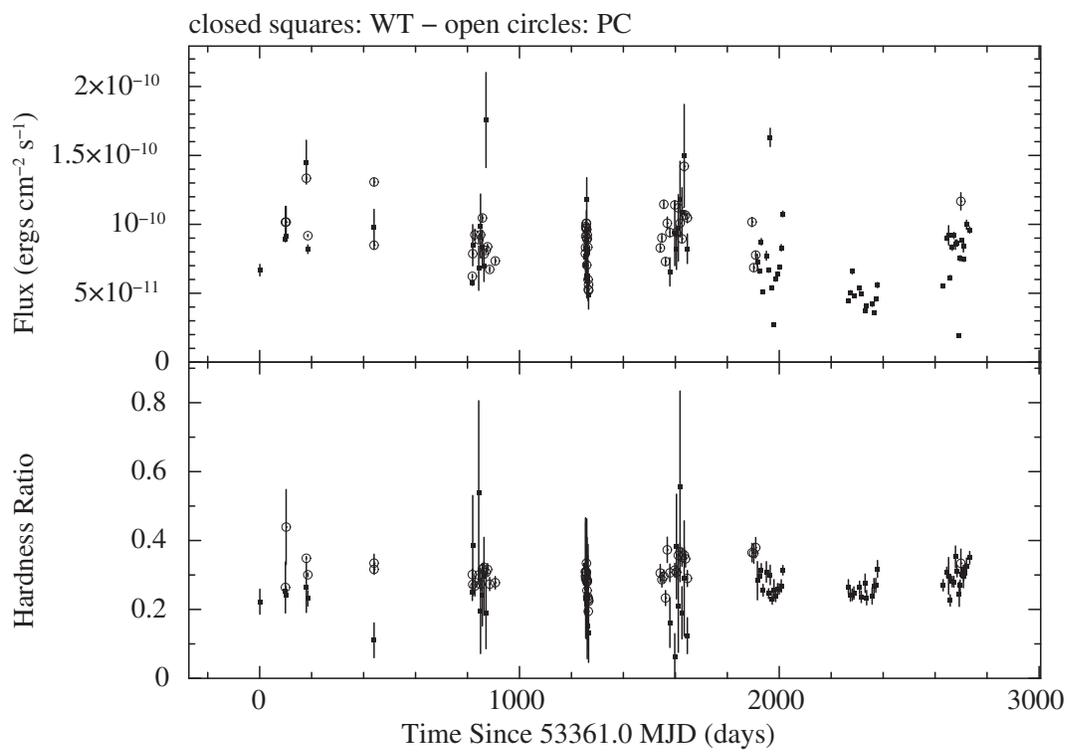}
\caption{Unabsorbed flux light curve and hardness ratio for H 1426+428.}
\label{fig:h_1426+428_light_curve}
\end{figure*}

\begin{figure*}
\includegraphics[angle=270,width=15cm]{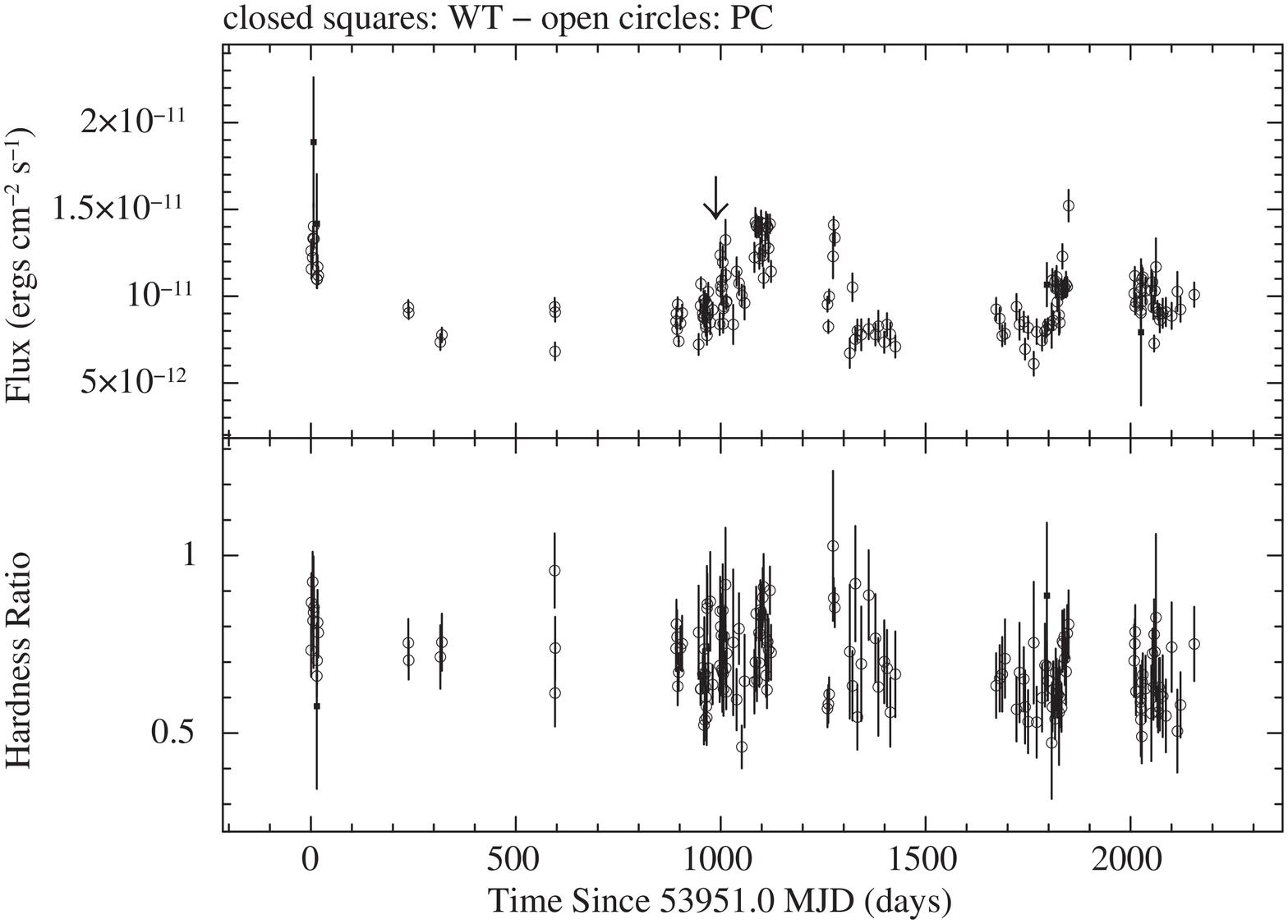}
\caption{Unabsorbed flux light curve and hardness ratio for PKS 1510-089. Three sigma upper limits are denoted by downward pointing arrows.}
\label{fig:pks_1510-089_light_curve}
\end{figure*}

\begin{figure*}
\includegraphics[angle=270,width=15cm]{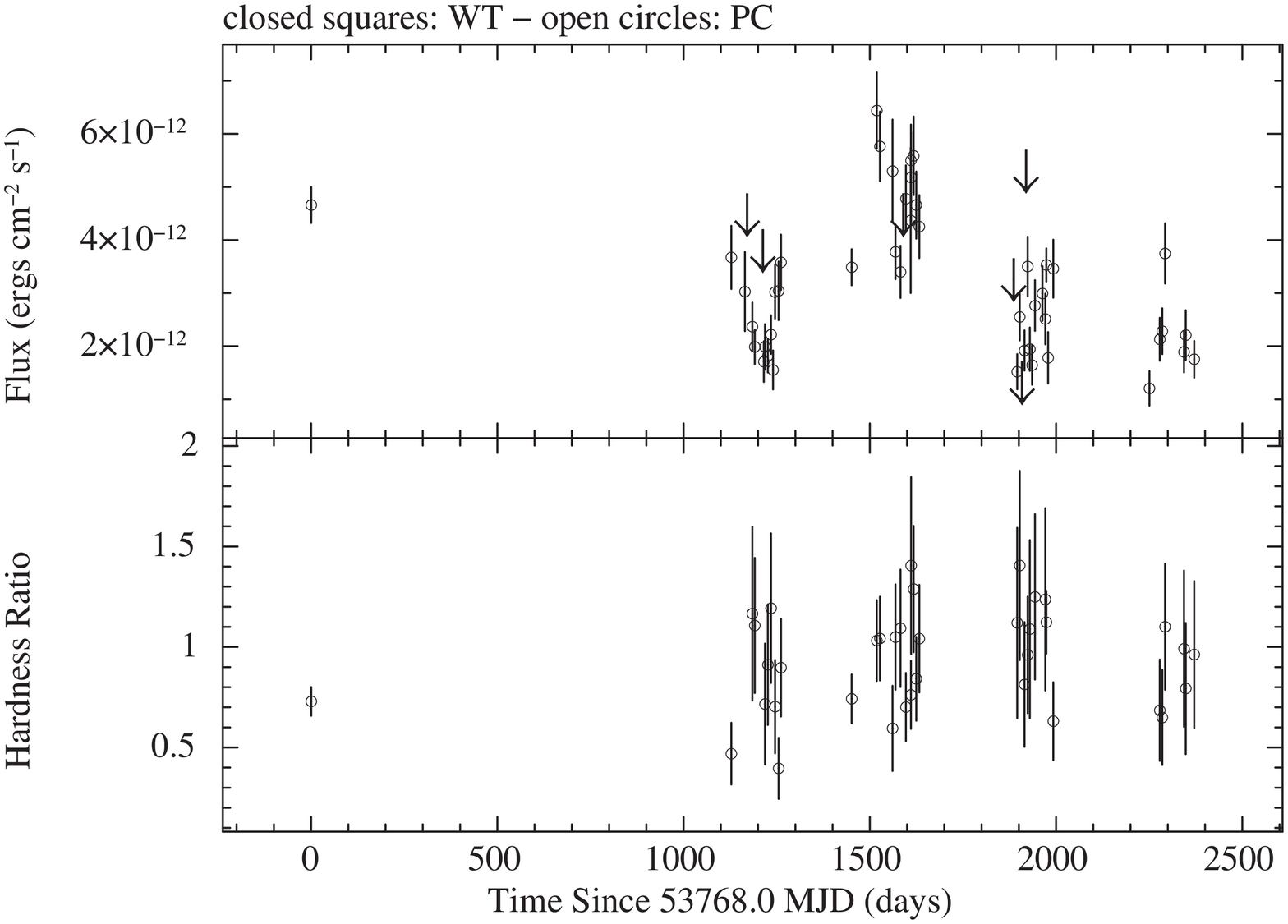}
\caption{Unabsorbed flux light curve and hardness ratio for PKS 1622-297. Three sigma upper limits are denoted by downward pointing arrows.}
\label{fig:pks_1622-297_light_curve}
\end{figure*}

\begin{figure*}
\includegraphics[angle=270,width=15cm]{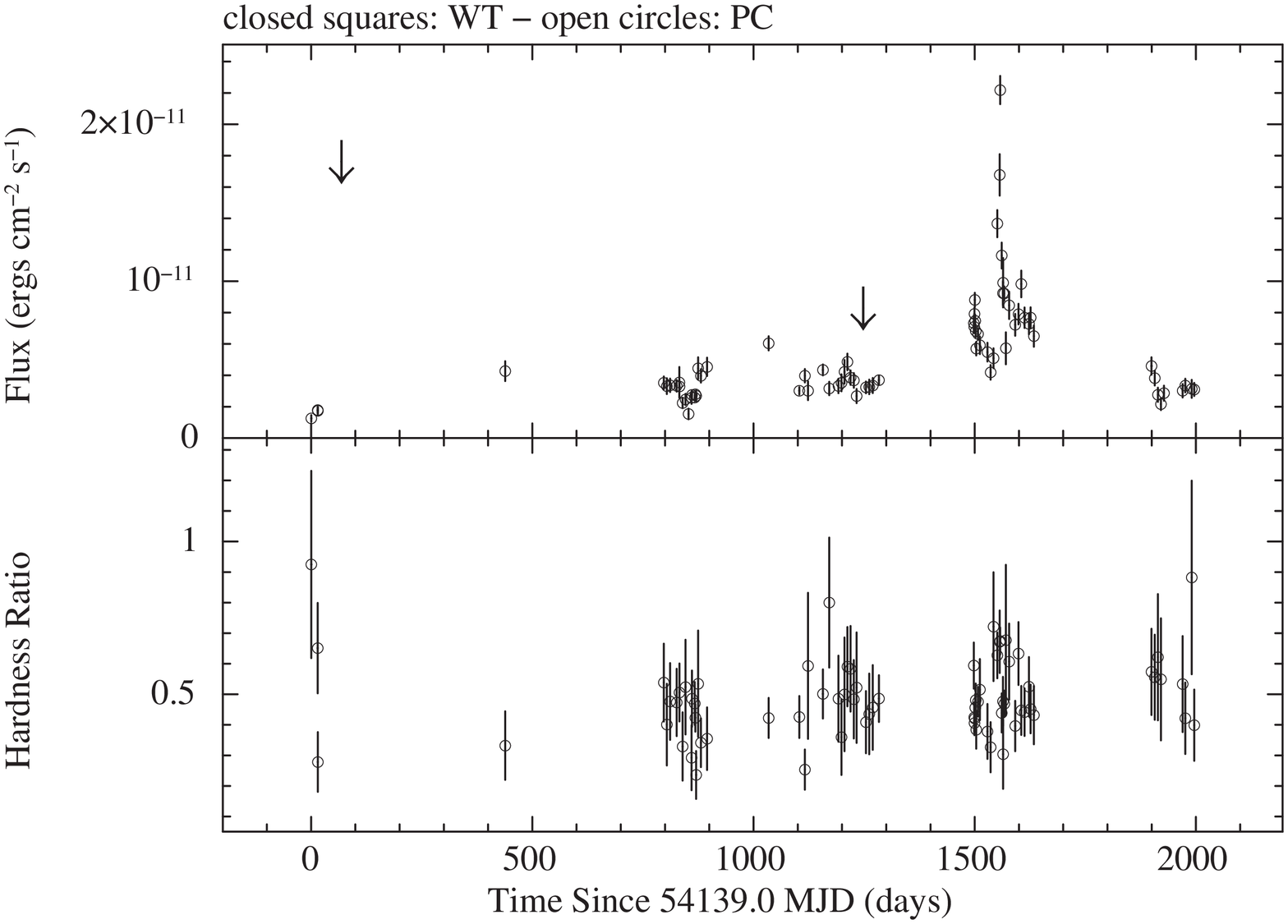}
\caption{Unabsorbed flux light curve and hardness ratio for 1Jy 1633+38. Three sigma upper limits are denoted by downward pointing arrows.}
\label{fig:1jy_1633+38_light_curve}
\end{figure*}

\begin{figure*}
\includegraphics[angle=270,width=15cm]{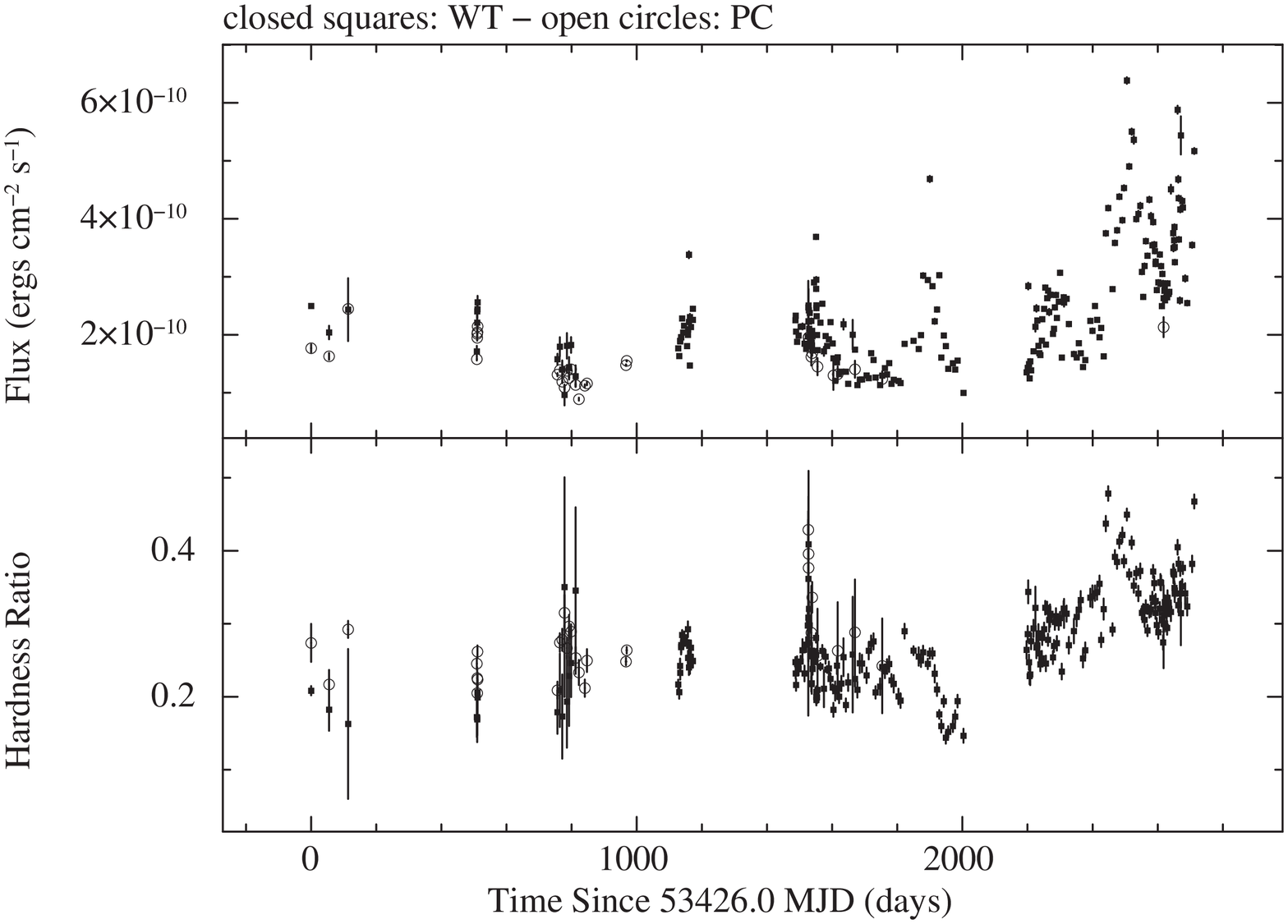}
\caption{Unabsorbed flux light curve and hardness ratio for Mrk 501.}
\label{fig:mrk_501_light_curve}
\end{figure*}

\begin{figure*}
\includegraphics[angle=270,width=15cm]{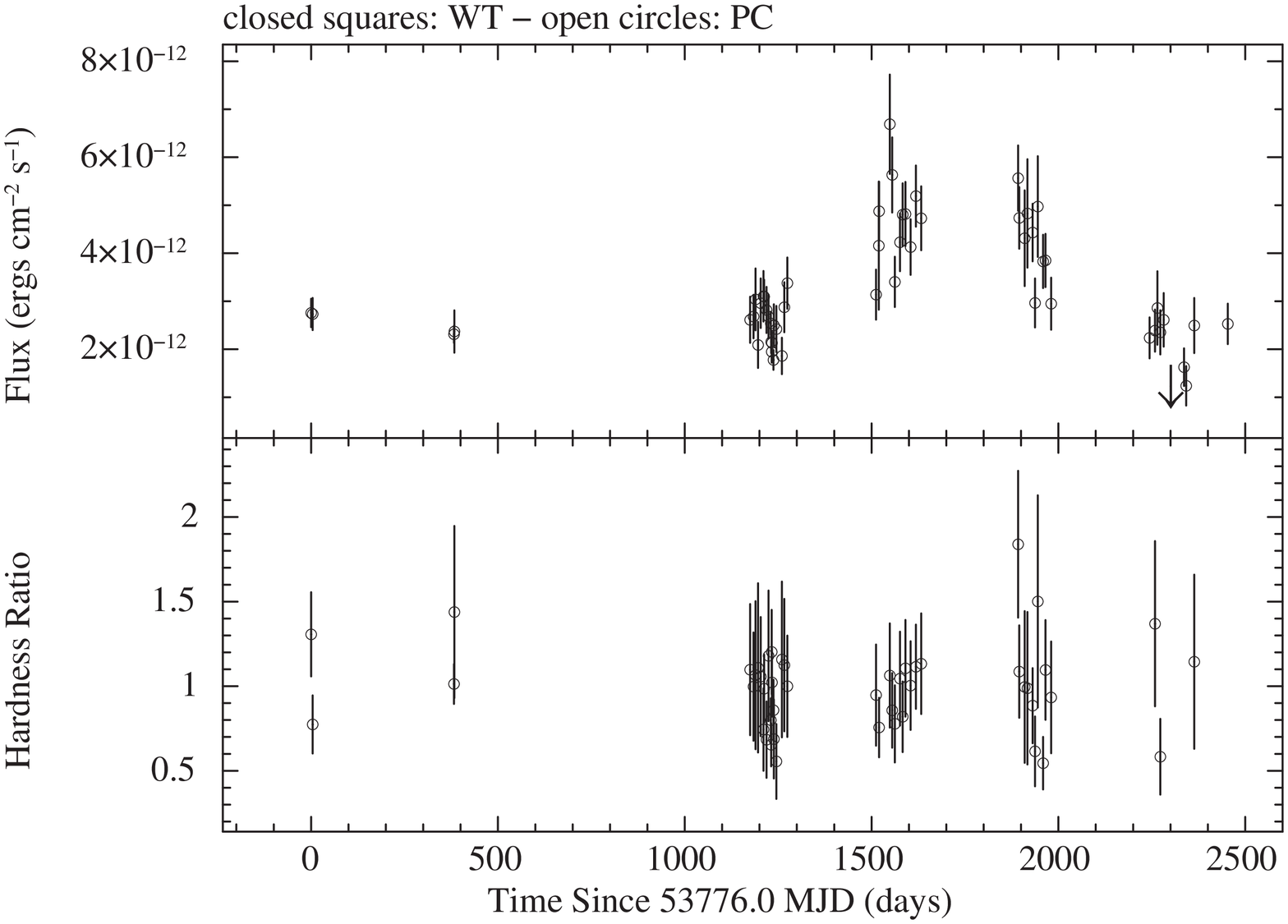}
\caption{Unabsorbed flux light curve and hardness ratio for PKS 1730-130. Three sigma upper limits are denoted by downward pointing arrows.}
\label{fig:pks_1730-130_light_curve}
\end{figure*}

\begin{figure*}
\includegraphics[angle=270,width=15cm]{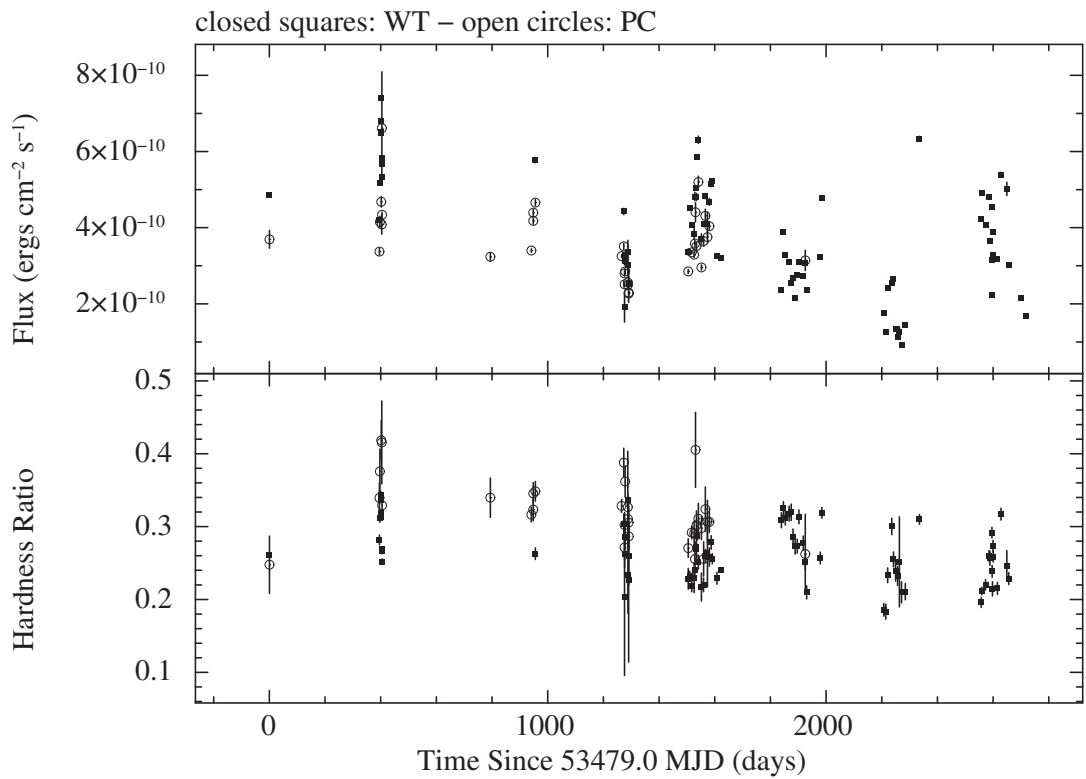}
\caption{Unabsorbed flux light curve and hardness ratio for 1ES 1959+650.}
\label{fig:1es_1959+650_light_curve}
\end{figure*}

\begin{figure*}
\includegraphics[angle=270,width=15cm]{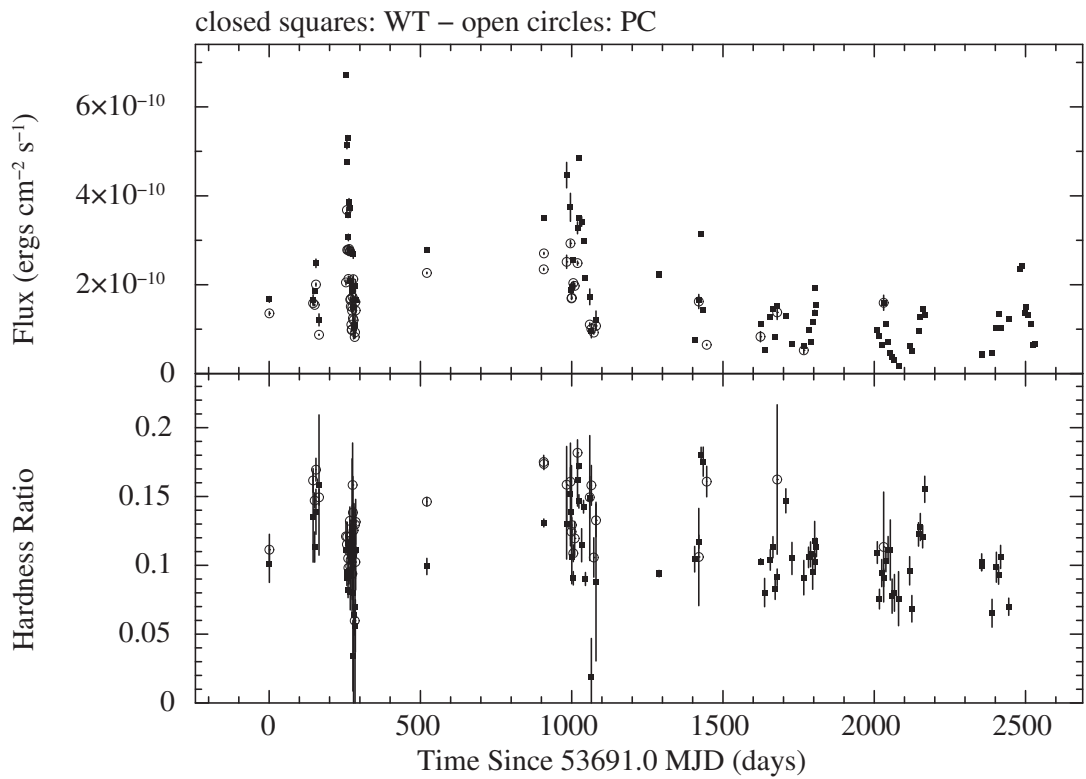}
\caption{Unabsorbed flux light curve and hardness ratio for PKS 2155-304.}
\label{fig:pks_2155-304_light_curve}
\end{figure*}

\begin{figure*}
\includegraphics[angle=270,width=15cm]{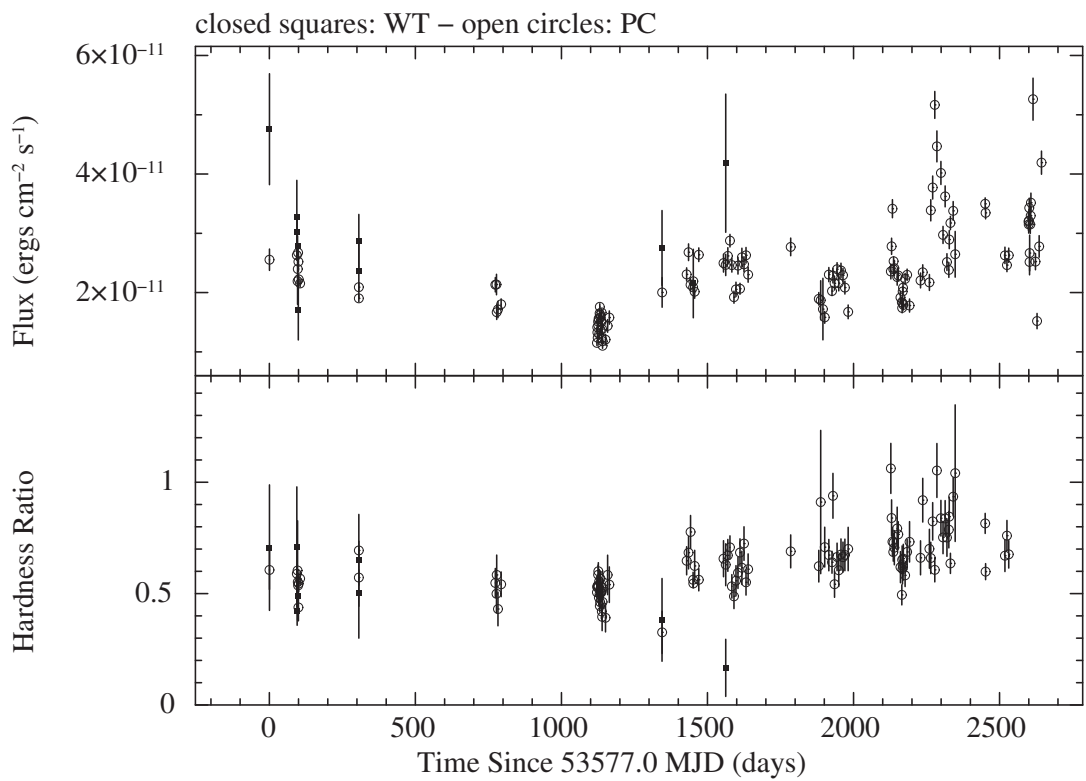}
\caption{Unabsorbed flux light curve and hardness ratio for BL Lacertae.}
\label{fig:bl_lacertae_light_curve}
\end{figure*}

\begin{figure*}
\includegraphics[angle=270,width=15cm]{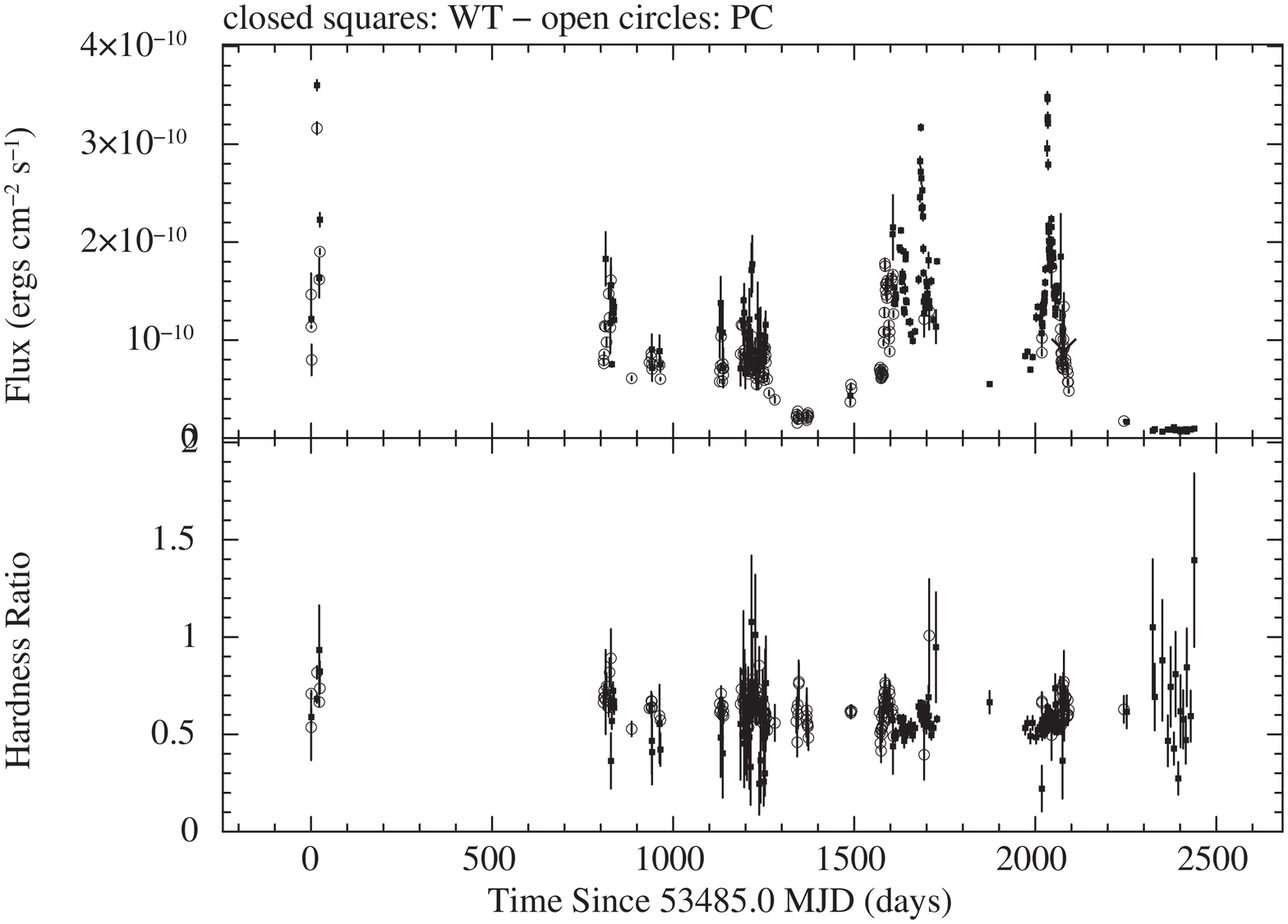}
\caption{Unabsorbed flux light curve and hardness ratio for 3C 454.3. Three sigma upper limits are denoted by downward pointing arrows.}
\label{fig:3c_454p3_light_curve}
\end{figure*}

\begin{figure*}
\includegraphics[angle=270,width=15cm]{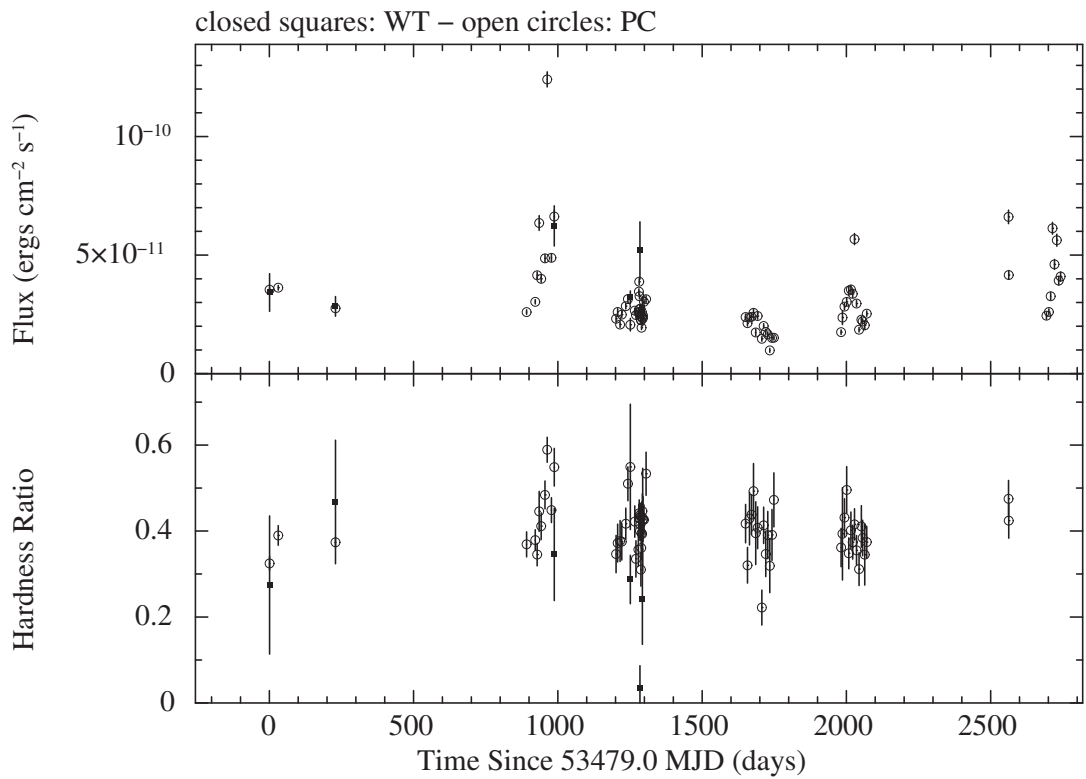}
\caption{Unabsorbed flux light curve and hardness ratio for 1ES 2344+514.}
\label{fig:1es_2344+514_light_curve}
\end{figure*}

We applied the count rate to unabsorbed flux conversion factors calculated in section~\ref{sec:spectralFitting} to the overall light curves to create source flux light curves which are presented alongside their hardness ratios in figures \ref{fig:pks_0208-512_light_curve} - \ref{fig:1es_2344+514_light_curve}. The source flux light curves use an x-axis in units of days and are offset in MJD by $T_0$ where $T_0 \equiv \lfloor Min(x_i-\delta_i)\rfloor$ and $x_i$ is the center of the time bin and $\delta_i$ is the half width of the bin. 

There are three month gaps in most of the data sets due to the fact that most of the sources enter \emph{Swift}'s Sun constraint at some point during the year. Some gaps are even longer due to the \emph{Swift} team's goal of maximizing observing time in the part of the sky opposite from the Sun, in order to encourage ground based follow-up of gamma ray bursts detected by \emph{Swift}. When targets are outside this area, they are less likely to receive observations.

A selection of sources whose light curves have interesting or remarkable features are discussed in the sections below. For some sources, the error bars were large enough or the sampling not long enough to say anything definitive about flaring periods using these data alone.

\subsubsection{PKS 0208-512}
Figure \ref{fig:pks_0208-512_light_curve} shows the light curve and hardness ratio of PKS 0208-512 from 2005 April 1 through 2011 December 30. Observations from MJD 54709 to 54887 show the source brightening from $2\times10^{-12}$ up to $4\times10^{-12}~ergs~cm^{-2}~s^{-1}$ before returning down to $2\times10^{-12}~ergs~cm^{-2}~s^{-1}$. From MJD 55084 to 55203, the source maintained a flux around $2\times10^{-12}~ergs~cm^{-2}~s^{-1}$, but the count rate from MJD 55460 to 55559 indicated an extended mild high state, with flux levels from $3\times10^{-12}$ to $4\times10^{-12}~ergs~cm^{-2}~s^{-1}$. Observations from MJD 55756 to 55926 indicate that the flux was decreasing from $\approx3\times10^{-12}$ down to $\approx10^{-12}~ergs~cm^{-2}~s^{-1}$.

\subsubsection{PKS 0235+164}
Figure \ref{fig:pks_0235+164_light_curve} shows the light curve and hardness ratio of PKS 0235+164 from 2005 June 28 through 2012 January 26. The source flux is measured to be between $3\times10^{-12}~ergs~cm^{-2}~s^{-1}$ and $2\times10^{-11}~ergs~cm^{-2}~s^{-1}$ for the majority of the observations, but it exhibited a very strong high state on MJD 54761 when the flux increased to $(9.0 \pm 0.3) \times 10^{-11}~ergs~cm^{-2}~s^{-1}$. During and following this high flux state, a softening in the spectrum (hardness ratio $<0.5$) was measured from MJD 54746 through MJD 54818, although the flux level had already decreased during part of this time period.

\subsubsection{LS I +61 303}
Figure \ref{fig:lsi+61_303_light_curve} shows the light curve and hardness ratio of LS I+61 303, the lone X-ray/TeV binary in our sample, from 2006 September 2 through 2012 February 5. Based on prior studies (e.g. \citet{smith09}), it is not surprising that it shows regular variability, with a flux usually between $10^{-11}$ and $3\times10^{-11}~ergs~cm^{-2}~s^{-1}$, with occasional variability outside of those limits.  Variability on timescales significantly less than the 26.5 day orbital period is frequently observed.  The hardness ratio also shows regular variability, with values ranging from 1 to 2.

\subsubsection{PKS 0528+134}
Figure \ref{fig:pks_0528+134_light_curve} shows the light curve and hardness ratio of PKS 0528+134 from 2006 March 28 through 2012 February 26. This source exhibited variability on multiple timescales throughout most of the long-term lightcurve. Observations from MJD 54732 to 55619 regularly indicate a source flux below $5 \times 10^{-12}~ergs~cm^{-2}~s^{-1}$; however, the source flux tended to be greater than $5 \times 10^{-12}~ergs~cm^{-2}~s^{-1}$ from MJD 53822 to 53847 and MJD 55810 to 55983.

\subsubsection{S5 0716+714}
Figure \ref{fig:s5_0716+714_light_curve} shows the light curve and hardness ratio of S5 0716+714 from 2005 April 2 through 2012 March 25. The light curve is well sampled and shows strong variability from MJD 54375 to 54589 and MJD 55120 to 55285 and shows a peak flux of $(7.7 \pm 1.2) \times 10^{-11}~ergs~cm^{-2}~s^{-1}$ on MJD 54586.

\subsubsection{QSO B0827+243}
Figure \ref{fig:qso_b0827+243_light_curve} shows the light curve and hardness ratio of QSO B0827+243 from 2008 January 11 through 2012 March 31. This light curve received less sampling than some of the others, but it does show some variability, especially at MJD 55503, when the flux is at a maximum of $(9.6 \pm 0.5) \times 10^{-12}~ergs~cm^{-2}~s^{-1}$ and decreases in brightness over the next sixty days to near its average of $(2.93 \pm 0.10) \times 10^{-12}~ergs~cm^{-2}~s^{-1}$

\subsubsection{OJ 287}
Figure \ref{fig:oj_287_light_curve} shows the light curve and hardness ratio of OJ 287 from 2005 May 20 through 2012 April 23. Observations prior to MJD 54455 show the flux below $7 \times 10^{-12}~ergs~cm^{-2}~s^{-1}$, but the observations since MJD 54763 exhibit more variability in the light curve, with fluxes ranging from $(3.4 \pm 0.4) \times 10^{-12}$ up to $(1.74 \pm 0.10) \times 10^{-11}~ergs~cm^{-2}~s^{-1}$.

\subsubsection{Mrk 421}
Figure \ref{fig:mrk_421_light_curve} shows the light curve and hardness ratio of Mrk 421 from 2005 March 1 through 2012 May 31. In addition to being the most often observed source in our sample, Mrk 421 is the brightest source in our sample with a flux ranging from $(8.0 \pm 0.2) \times 10^{-11}$ to $(6.521 \pm 0.009) \times 10^{-9}~ergs~cm^{-2}~s^{-1}$ and with a mean flux of $(9.049 \pm 0.007) \times 10^{-10}~ergs~cm^{-2}~s^{-1}$. Every year that the source has been observed, there have been observations where the flux is $< 2 \times 10^{-9}~ergs~cm^{-2}~s^{-1}$, but the time periods from MJD 53565 to 53931, MJD 54380 to 54779, and MJD 55147 to 55392 show high flux and variability. The hardness ratio also indicates enhanced spectral variability during the time period when the flux is most variable.  The source has been weaker and less variable since MJD 55533, with a typical flux $< 2 \times 10^{-9}~ergs~cm^{-2}~s^{-1}$ and a softening in the spectrum with a hardness ratio typically $<0.2$.

\subsubsection{W Com}
Figure \ref{fig:w_com_light_curve} shows the light curve and hardness ratio of W Com from 2005 July 14 through 2012 May 25. As noted in section \ref{sec:excessVariance}, this source has been well sampled and shows significant variability, with flux ranging from $1\times 10^{-12}$ to $\sim3\times 10^{-11}~ergs~cm^{-2}~s^{-1}$.  The time period from MJD 55599 to 56073, which extends to the end of our sample, seems to indicate the source is in a nearly quiescent state with the source flux consistently $< 4\times 10^{-12}~ergs~cm^{-2}~s^{-1}$.

\subsubsection{PKS 1622-297}
Figure \ref{fig:pks_1622-297_light_curve} shows the light curve and hardness ratio of PKS 1622-297 from 2006 February 2 through 2012 July 30. The majority of the observations of this source have a flux between $1\times10^{-12}$ and $4\times 10^{-12}~ergs~cm^{-2}~s^{-1}$, but during the observations from MJD 55219 to 55401, the source brightened to $3\times 10^{-12}$ to $\sim6\times 10^{-12}~ergs~cm^{-2}~s^{-1}$.

\subsubsection{1Jy 1633+38}
Figure \ref{fig:1jy_1633+38_light_curve} shows the light curve and hardness ratio of 1Jy 1633+38 from 2007 February 8 through 2012 July 27. Normally this source has a flux $< 5\times 10^{-12}~ergs~cm^{-2}~s^{-1}$, but the observations from MJD 55636 to 55772 showed a brightening, with the flux spiking up to $(2.19 \pm 0.09) \times 10^{-11}~ergs~cm^{-2}~s^{-1}$ on MJD 55696.

\subsubsection{Mrk 501}
Figure \ref{fig:mrk_501_light_curve} shows the light curve and hardness ratio of Mrk 501 from 2005 February 25 through 2012 July 30. Since Mrk 501 never enters the Sun constraint for \emph{Swift}, the target doesn't have as many large coverage gaps as the other sources, which is demonstrated in the periods from MJD 54913 to 55429 and MJD 55623 to 56138. The flux of the source tended to stay between $\approx1 \times 10^{-10}$ and $3 \times 10^{-10}~ergs~cm^{-2}~s^{-1}$, but after MJD 55866, it has brightened to values between $3 \times 10^{-10}$ and $6 \times 10^{-10}~ergs~cm^{-2}~s^{-1}$. During this same brightening period the spectrum has also hardened, with the hardness ratio $\approx0.35$, whereas its typical average during the early part of the light curve was $\approx 0.25$.

\subsubsection{PKS 2155-304}
Figure \ref{fig:pks_2155-304_light_curve} shows the light curve and hardness ratio of PKS 2155-304 from 2005 November 17 through 2012 July 27. From MJD 53691 to 54771, the light curve is highly variable, with multiple high states including one that reaches nearly $7\times10^{-10}~ergs~cm^{-2}~s^{-1}$.  From MJD 54979 to 56135, significant variability continued, but the flux only reached $\approx3\times10^{-10}~ergs~cm^{-2}~s^{-1}$ and for the most part remained below $2\times10^{-11}~ergs~cm^{-2}~s^{-1}$.

\subsubsection{3C 454.3}
Figure \ref{fig:3c_454p3_light_curve} shows the light curve and hardness ratio of 3C 454.3 from 2005 April 24 through 2011 December 29. It is characterized by extended high states ($>months$) with superimposed shorter timescale flares.  During each year of monitoring, the flux can span almost an entire order of magnitude from $\approx5\times10^{-11}$ up to $\approx4\times10^{-10}~ergs~cm^{-2}~s^{-1}$. Observations from MJD 54749 to 54978 show a decrease in the flux rate down to $\approx2 \times 10^{-11}~ergs~cm^{-2}~s^{-1}$. The next notable period of decreased flux is from MJD 55573 to 55924, which is at the end of our sample where the source stayed in quiescence with a flux at $\approx1 \times 10^{-11}~ergs~cm^{-2}~s^{-1}$.

\subsubsection{1ES 2344+514}
Figure \ref{fig:1es_2344+514_light_curve} shows the light curve and hardness ratio of 1ES 2344+514 from 2005 April 19 through 2012 April 24. The flux of this source only averages $(2.63 \pm 0.03) \times 10^{-11}~ergs~cm^{-2}~s^{-1}$, but on MJD 54442 its flux reached up to $(1.24 \pm 0.03) \times 10^{-10}~ergs~cm^{-2}~s^{-1}$.

\subsection{Online Repository}
\label{sec:onlineRepository}
In addition to the data products in this paper, we present light curve and hardness ratio text files and plots at \url{www.swift.psu.edu/monitoring/}. In addition to the time and rate information, the standard light curve text files also contain the fractional exposure time and the mode information. More detailed text files are also produced that contain the non-background subtracted rate in the source region, non-PSF corrected rate in the source region, and the background count rates. Tables \ref{lcFields} and \ref{hrFields} summarize what can be found in the light curve and hardness ratio text files respectively. Light curve and hardness ratio text files and plots are produced for the complete \emph{Swift}-XRT data set, with detailed fine-binning for single observations. Since the online data products are updated several times a day, they include observations on these sources from recent data beyond the dates reported in this paper.

The monitoring web site also contains data on over 100 additional sources of interest to the high energy astrophysical community, 121 of which are listed with their total exposure times in table \ref{publicSources}. Most of these sources were added to the Swift automated monitoring list after a report of a high state, which typically originated from \emph{Fermi}-LAT observations. The data set for this expanded list was obtained from 2004 December 17 through 2012 August 31.

\section{Summary}
We have presented a summary of our analysis of the \emph{Swift}-XRT observations of the 23 \emph{Fermi}-LAT sources of interest through August 31st, 2012. We have also described the publicly-available and near-real-time data analysis that can be found in an online repository.  Light curves and hardness ratio curves are available. The real-time availability of these data provide a means for observers to monitor the activity of sources for existing campaigns and to trigger target-of-opportunity observations when these sources are seen to be in an interesting state.

As expected, it is clear from the light curves that variability is present on multiple timescales.  The excess variance demonstrated that the sources are just as variable on longer time scales as on shorter ones. The light curves show some sources with moderate to low activity, punctuated with large high states, and other sources with nearly continual variability.  Variability of approximately two orders of magnitude is observed.  After an inspection for any possible systematic effects, these data can be compared to data from other observatories to search for both long term and short term correlations and to produce broad band spectral energy distributions.

\acknowledgements
We gratefully acknowledge the \emph{Swift} Team for making these observations possible, and we are grateful for support from NASA grants NNX10AU14G and NNX09AU07G.

{\it Facilities:} \facility{Swift (XRT)}

\begin{deluxetable}{rrrr}
\tabletypesize{\scriptsize}
\tablecaption{Photon Counting region definitions\label{pcregions}}
\tablewidth{0pt}
\tablecolumns{4}
\tablehead{
\colhead{Minimum Count Rate} & 
\colhead{Maximum Count Rate} & 
\colhead{Source Region Radius} & 
\colhead{Background Region Radii} \\
\colhead{($counts \enskip s^{-1}$)} &
\colhead{($counts \enskip s^{-1}$)} &
\colhead{($\arcsec$)} &
\colhead{($\arcsec$)}
}
\startdata
  0.000 & 0.001 & 12 & 40 - 140 \\
  0.001 & 0.005 & 17 & 40 - 140 \\
  0.005 & 0.010   & 22 & 45 - 145 \\
  0.010 & 0.050   & 25 & 50 - 150 \\
  0.050 & 0.100   & 45 & 90 - 190 \\
  0.100 & 0.500     & 55 & 110 - 210 \\
  0.500 & 1.000   & 5 - 65 & 130 - 230 \\
  1.000 & 3.000   & 10 - 75 & 150 - 250 \\
  3.000 & 5.000   & 12 - 85 & 170 - 270 \\
  5.000 & 7.000   & 15 - 95 & 190 - 290 \\
  7.000 & $\infty$ & 20 - 105 & 210 - 310 \\
\enddata
\tablecomments{Minimum count rates are exclusive while maximum count rates are inclusive. If the count rate is $>0.5\enskip counts \enskip s^{-1}$, source region is an annulus.}
\end{deluxetable}

\begin{deluxetable}{rrrrr}
\tabletypesize{\scriptsize}
\tablecaption{Windowed Timing region definitions\label{wtregions}}
\tablewidth{0pt}
\tablecolumns{5}
\tablehead{
\colhead{Minimum Count Rate} & 
\colhead{Maximum Count Rate} & 
\colhead{Source Region Width} & 
\colhead{Pileup Region Width} & 
\colhead{Background Region Width} \\
\colhead{($counts \enskip s^{-1}$)} &
\colhead{($counts \enskip s^{-1}$)} &
\colhead{($\arcsec$)} &
\colhead{($\arcsec$)} &
\colhead{($\arcsec$)}
}
\startdata
  0.       & 1.        & 35   & None & 150 - 450 \\
  1.       & 5.        & 71   & None & 200 - 450 \\
  5.       & 10.      & 118 & None & 260 - 450 \\
  10.     & 50.      & 165 & None & 300 - 450 \\
  50.     & 100.    & 200 & None & 450 - 500 \\
  100.   & 300.    & 236 & None & 450 - 550 \\
  300.   & 400.    & 300 & 9.43 & 450 - 640 \\
  400.   & 600.    & 450 & 23.65 & 450 - 1200 \\
  600.   & 1000.    & 450 & 28.38 & 450 - 1200 \\
  1000.   & $\infty$  & 450 & 37.84 & 450 - 1200 \\
\enddata
\tablecomments{Minimum count rates are exclusive while maximum count rates are inclusive. If the count rate is $>300\enskip counts \enskip s^{-1}$, a pileup region is subtracted from the source region.}
\end{deluxetable}

\begin{deluxetable}{lrrrrr}
\tabletypesize{\scriptsize}
\tablecaption{Spectral Fitting Parameters\label{specFitParameters}}
\tablewidth{0pt}
\tablecolumns{6}
\tablehead{
\colhead{Source Name} & 
\colhead{$\langle N_H \rangle$} &
\colhead{$\langle \alpha \rangle$} & 
\colhead{$\langle c \rangle$} &
\colhead{$\langle\frac{\chi^{2}}{d.o.f}\rangle$} &
\colhead{Total Degrees} \\
\colhead{} & 
\colhead{($cm^{-2}$)} & 
\colhead{} & 
\colhead{($photons~cm^{-2}~s^{-1}$)} &
\colhead{} &
\colhead{of Freedom}
}
\startdata
PKS 0208-512 & $ (5^{+4}_{-3}) \times 10^{20} $ & $ 1.83^{+0.14}_{-0.13} $ & $ (4.2^{+0.6}_{-0.5}) \times 10^{-4} $ & 0.89 & 265 \\
PKS 0235+164 & $ (2.7^{+0.3}_{-0.2}) \times 10^{21} $ & $ 1.93^{+0.07}_{-0.07} $ & $ (9.9^{+0.7}_{-0.7}) \times 10^{-4} $ & 0.99 & 530 \\
LS I +61 303 & $ (7.2^{+0.6}_{-0.6}) \times 10^{21} $ & $ 1.71^{+0.07}_{-0.07} $ & $ (2.7^{+0.3}_{-0.2}) \times 10^{-3} $ & 1.07 & 1029 \\
PKS 0528+134 & $ (5.4^{+0.8}_{-0.8}) \times 10^{21} $ & $ 1.75^{+0.12}_{-0.12} $ & $ (7.6^{+1.3}_{-1.1}) \times 10^{-4} $ & 1.06 & 287 \\
S5 0716+714 & $ (1.6^{+0.8}_{-0.8}) \times 10^{20} $ & $ 2.31^{+0.06}_{-0.06} $ & $ (2.07^{+0.07}_{-0.07}) \times 10^{-3} $ & 1.19 & 587 \\
QSO B0827+243 & $ (7 ^{+2}_{-2}) \times 10^{20} $ & $ 1.80^{+0.09}_{-0.08} $ & $ (4.6^{+0.3}_{-0.3}) \times 10^{-4} $ & 1.09 & 187 \\
OJ 287 & $ (9^{+8}_{-6}) \times 10^{20} $ & $ 2.0^{+0.3}_{-0.2} $ & $ (7.1^{+1.3}_{-0.9}) \times 10^{-4} $ & 0.89 & 488 \\
Mrk 421 & $ (1.10^{+0.06}_{-0.06}) \times 10^{21} $ & $ 2.322^{+0.014}_{-0.013} $ & $ (3.53^{+0.05}_{-0.05}) \times 10^{-1} $ & 1.38 & 21191\\
W Com & $ (2.9^{+1.4}_{-1.7}) \times 10^{20} $ & $ 2.83^{+0.14}_{-0.10} $ & $ (7.1^{+0.5}_{-0.4}) \times 10^{-4} $ & 1.07 & 323 \\
3C 273 & $ (2.3^{+0.5}_{-1.5}) \times 10^{20} $ & $ 1.62^{+0.02}_{-0.02} $ & $ (2.25^{+0.03}_{-0.03}) \times 10^{-2} $ & 1.23 & 2988 \\
3C 279 & $ (5.3^{+0.7}_{-0.7}) \times 10^{20} $ & $ 1.70^{+0.03}_{-0.03} $ & $ (2.17^{+0.06}_{-0.06}) \times 10^{-3} $ & 1.09 & 1424 \\
1Jy 1406-076 & $ (3^{+6}_{-3}) \times 10^{21} $ & $ 3^{+1.6}_{-1.0} $ & $ (1.6^{+6.6}_{-1.3}) \times 10^{-4} $ & 1.09 & 38 \\
H 1426+428 & $ (5.2^{+1.5}_{-1.1}) \times 10^{20} $ & $ 2.06^{+0.06}_{-0.06} $ & $ (1.40^{+0.07}_{-0.06}) \times 10^{-2} $ & 1.19 & 2021 \\
PKS 1510-089 & $ (9^{+3}_{-3}) \times 10^{20} $ & $ 1.38^{+0.08}_{-0.08} $ & $ (1.09^{+0.10}_{-0.09}) \times 10^{-3} $ & 0.99 & 816 \\
PKS 1622-297 & $ (3^{+3}_{-3}) \times 10^{21} $ & $ 1.7^{+0.4}_{-0.4} $ & $ (6^{+4}_{-2}) \times 10^{-4} $ & 0.87 & 90 \\
1Jy 1633+38 & $ (3.5^{+12.4}_{-1.4}) \times 10^{20} $ & $ 1.5^{+0.5}_{-0.4} $ & $ (3.6^{+0.6}_{-0.4}) \times 10^{-4} $ & 1.10 & 297 \\
Mrk 501 & $ (5^{+3}_{-2}) \times 10^{20} $ & $ 2.18^{+0.09}_{-0.07} $ & $ (3.0^{+0.3}_{-0.2}) \times 10^{-2} $ & 1.12 & 1845 \\
PKS 1730-130 & $ (4^{+3}_{-2}) \times 10^{21} $ & $ 1.6^{+0.3}_{-0.2} $ & $ (3.4^{+1.4}_{-0.8}) \times 10^{-4} $ & 0.89 & 108 \\
1ES 1959+650 & $ (1.70^{+0.17}_{-0.15}) \times 10^{21} $ & $ 2.32^{+0.07}_{-0.06} $ & $ (7.4^{+0.4}_{-0.3}) \times 10^{-2} $ & 1.16 & 2015 \\
PKS 2155-304 & $ (3.8^{+0.5}_{-0.5}) \times 10^{20} $ & $ 2.66^{+0.04}_{-0.04} $ & $ (3.76^{+0.09}_{-0.08}) \times 10^{-2} $ & 1.25 & 2112 \\
BL Lacertae & $ (2.7^{+0.2}_{-0.2}) \times 10^{21} $ & $ 1.94^{+0.05}_{-0.05} $ & $ (3.4^{+0.2}_{-0.2}) \times 10^{-3} $ & 1.13 & 793 \\
3C 454.3 & $ (1.7^{+0.2}_{-0.2}) \times 10^{21} $ & $ 1.59^{+0.04}_{-0.04} $ & $ (1.39^{+0.08}_{-0.07}) \times 10^{-2} $ & 1.09 & 2584 \\
1ES 2344+514 & $ (1.98^{+0.12}_{-0.11}) \times 10^{21} $ & $ 2.24^{+0.05}_{-0.04} $ & $ (5.3^{+0.2}_{-0.2}) \times 10^{-3} $ & 1.10 & 717 \\
\enddata
\end{deluxetable}

\begin{deluxetable}{lrrrrrr}
\tabletypesize{\scriptsize}
\rotate
\tablecaption{Original LAT 23 Sources\label{sources}}
\tablewidth{0pt}
\tablecolumns{7}
\tablehead{
\colhead{Source Name} & 
\colhead{Total XRT} &
\colhead{Minimum Rate} & 
\colhead{Maximum Rate} &
\colhead{Mean Rate} & 
\colhead{$\sigma_{rms1}^2$} &
\colhead{$\sigma_{rms2}^2$} \\
\colhead{} & 
\colhead{Exposure Time (ks)} &
\colhead{($counts \enskip s^{-1}$)} & 
\colhead{($counts \enskip s^{-1}$)} &
\colhead{($counts \enskip s^{-1}$)} & 
\colhead{} &
\colhead{} 

}
\startdata
PKS 0208-512 & 139 & $(2.0 \pm 0.6) \times 10^{-2} $ & $(1.8 \pm 0.6) \times 10^{-1} $ & $(6.12 \pm 0.08) \times 10^{-2} $ & 0.12 & 0.06 \\
PKS 0235+164 & 187 & $(2.0 \pm 0.5) \times 10^{-2} $ & $ 1.13 \pm 0.04 $ & $(6.86 \pm 0.07) \times 10^{-2} $ & 1.89 & 1.23 \\
LS I +61 303 & 309 & $(2.1 \pm 0.6) \times 10^{-2} $ & $(5.3 \pm 0.6) \times 10^{-1} $ & $(1.616 \pm 0.008) \times 10^{-1} $ & 0.22 & 0.13 \\
PKS 0528+134 & 148 & $(1.1 \pm 0.2) \times 10^{-2} $ & $(1.90 \pm 0.08) \times 10^{-1} $ & $(4.28 \pm 0.06) \times 10^{-2} $ & 0.34 & 0.29 \\
S5 0716+714 & 196 & $(8.5 \pm 1.1) \times 10^{-2} $ & $ 1.8 \pm 0.3 $ & $(2.844 \pm 0.014) \times 10^{-1} $ & 0.59 & 0.40 \\
QSO B0827+243 & 84 & $(3.6 \pm 0.7) \times 10^{-2} $ & $(2.16 \pm 0.09) \times 10^{-1} $ & $(6.58 \pm 0.10) \times 10^{-2} $ & 0.18 & 0.26 \\
OJ 287 & 201 & $(5.1 \pm 0.5) \times 10^{-2} $ & $(3.8 \pm 0.2) \times 10^{-1} $ & $(1.447 \pm 0.009) \times 10^{-1} $ & 0.15 & 0.11 \\
Mrk 421 & 778 & $ 1.82 \pm 0.05 $ & $(1.488 \pm 0.002) \times 10^{2} $ & $(2.0648 \pm 0.0006) \times 10^{1} $ & 0.49 & 0.34 \\
W Com & 147 & $(1.7 \pm 0.5) \times 10^{-2} $ & $(6 \pm 2) \times 10^{-1} $ & $(7.49 \pm 0.08) \times 10^{-2} $ & 1.03 & 0.95 \\
3C 273 & 292 & $ 2.80 \pm 0.02 $ & $ 8.5 \pm 0.2 $ & $ 4.312 \pm 0.005 $ & 0.06 & 0.07 \\
3C 279 & 509 & $(2.0 \pm 0.2) \times 10^{-1} $ & $(6.9 \pm 1.0) \times 10^{-1} $ & $(3.351 \pm 0.009) \times 10^{-1} $ & 0.05 & 0.02 \\
1Jy 1406-076 & 93 & $(1.0 \pm 0.3) \times 10^{-2} $ & $(3.5 \pm 0.6) \times 10^{-2} $ & $(1.94 \pm 0.06) \times 10^{-2} $ & 0.00 & 0.00 \\
H 1426+428 & 172 & $(5.0 \pm 0.3) \times 10^{-1} $ & $ 4.5 \pm 0.9 $ & $ 1.906 \pm 0.005 $ & 0.08 & 0.08 \\
PKS 1510-089 & 328 & $(1.23 \pm 0.14) \times 10^{-1} $ & $(3.8 \pm 0.8) \times 10^{-1} $ & $(1.971 \pm 0.009) \times 10^{-1} $ & 0.04 & 0.03 \\
PKS 1622-297 & 61 & $(2.0 \pm 0.5) \times 10^{-2} $ & $(1.08 \pm 0.10) \times 10^{-1} $ & $(4.71 \pm 0.10) \times 10^{-2} $ & 0.15 & 0.10 \\
1Jy 1633+38 & 139 & $(2.5 \pm 0.4) \times 10^{-2} $ & $(4.46 \pm 0.14) \times 10^{-1} $ & $(7.77 \pm 0.09) \times 10^{-2} $ & 0.43 & 0.39 \\
Mrk 501 & 330 & $ 2.15 \pm 0.06 $ & $(1.542 \pm 0.013) \times 10^{1} $ & $ 4.828 \pm 0.005 $ & 0.18 & 0.15 \\
PKS 1730-130 & 88 & $(1.9 \pm 0.6) \times 10^{-2} $ & $(1.0 \pm 0.2) \times 10^{-1} $ & $(4.07 \pm 0.08) \times 10^{-2} $ & 0.10 & 0.11 \\
1ES 1959+650 & 128 & $ 1.68 \pm 0.04 $ & $(1.360 \pm 0.008) \times 10^{1} $ & $ 6.498 \pm 0.008 $ & 0.12 & 0.09 \\
PKS 2155-304 & 204 & $(3.8 \pm 0.2) \times 10^{-1} $ & $(1.455 \pm 0.006) \times 10^{1} $ & $ 2.333 \pm 0.004 $ & 0.39 & 0.51 \\
BL Lacertae & 251 & $(1.83 \pm 0.07) \times 10^{-1} $ & $(8.6 \pm 0.4) \times 10^{-1} $ & $(3.040 \pm 0.013) \times 10^{-1} $ & 0.09 & 0.06 \\
3C 454.3 & 515 & $(9 \pm 2) \times 10^{-2} $ & $ 4.91 \pm 0.07 $ & $ 1.022 \pm 0.002 $ & 0.29 & 0.30 \\
1ES 2344+514 & 115 & $(1.69 \pm 0.14) \times 10^{-1} $ & $ 2.13 \pm 0.05 $ & $(4.51 \pm 0.02) \times 10^{-1} $ & 0.27 & 0.32 \\
\enddata
\tablecomments{Values are calculated for data through August 31st, 2012. $\sigma_{rms1}$ and $\sigma_{rms2}$ are calculated using the bins from the overall and finely binned light curves respectively. Sources are ordered by increasing R.A..}
\end{deluxetable}

\begin{deluxetable}{lrrrrrr}
\tabletypesize{\scriptsize}
\rotate
\tablecaption{Original LAT 23 Sources\label{sourcesFlux}}
\tablewidth{0pt}
\tablecolumns{5}
\tablehead{
\colhead{Source Name} & 
\colhead{Conversion Factor} & 
\colhead{Minimum Flux} & 
\colhead{Maximum Flux} &
\colhead{Mean Flux} \\
\colhead{} & 
\colhead{($ergs \enskip cm^{-2} \enskip counts^{-1}$)} &
\colhead{($ergs \enskip cm^{-2} \enskip s^{-1}$)} & 
\colhead{($ergs \enskip cm^{-2} \enskip s^{-1}$)} &
\colhead{($ergs \enskip cm^{-2} \enskip s^{-1}$)} & 

}
\startdata
PKS 0208-512 & $(3.75 \pm 0.12) \times 10^{-11} $ & $(7 \pm 2) \times 10^{-13} $ & $(7 \pm 2) \times 10^{-12} $ & $(2.30 \pm 0.08) \times 10^{-12} $\\
PKS 0235+164 & $(7.95 \pm 0.14) \times 10^{-11} $ & $(1.6 \pm 0.4) \times 10^{-12} $ & $(9.0 \pm 0.3) \times 10^{-11} $ & $(5.46 \pm 0.11) \times 10^{-12} $\\
LS I +61 303 & $(9.50 \pm 0.10) \times 10^{-11} $ & $(2.0 \pm 0.5) \times 10^{-12} $ & $(5.0 \pm 0.5) \times 10^{-11} $ & $(1.53 \pm 0.02) \times 10^{-11} $\\
PKS 0528+134 & $(8.2 \pm 0.2) \times 10^{-11} $ & $(9 \pm 2) \times 10^{-13} $ & $(1.56 \pm 0.08) \times 10^{-11} $ & $(3.52 \pm 0.11) \times 10^{-12} $\\
S5 0716+714 & $(4.21 \pm 0.04) \times 10^{-11} $ & $(3.6 \pm 0.5) \times 10^{-12} $ & $(7.7 \pm 1.2) \times 10^{-11} $ & $(1.199 \pm 0.012) \times 10^{-11} $\\
QSO B0827+243 & $(4.46 \pm 0.13) \times 10^{-11} $ & $(1.6 \pm 0.3) \times 10^{-12} $ & $(9.6 \pm 0.5) \times 10^{-12} $ & $(2.93 \pm 0.10) \times 10^{-12} $\\
OJ 287 & $(4.52 \pm 0.06) \times 10^{-11} $ & $(2.3 \pm 0.2) \times 10^{-12} $ & $(1.74 \pm 0.10) \times 10^{-11} $ & $(6.54 \pm 0.09) \times 10^{-12} $\\
Mrk 421 & $(4.383 \pm 0.003) \times 10^{-11} $ & $(8.0 \pm 0.2) \times 10^{-11} $ & $(6.521 \pm 0.009) \times 10^{-9} $ & $(9.049 \pm 0.007) \times 10^{-10} $\\
W Com & $(4.74 \pm 0.11) \times 10^{-11} $ & $(8 \pm 3) \times 10^{-13} $ & $(3.0 \pm 0.9) \times 10^{-11} $ & $(3.55 \pm 0.09) \times 10^{-12} $\\
3C 273 & $(4.273 \pm 0.014) \times 10^{-11} $ & $(1.197 \pm 0.011) \times 10^{-10} $ & $(3.62 \pm 0.10) \times 10^{-10} $ & $(1.843 \pm 0.006) \times 10^{-10} $\\
3C 279 & $(4.22 \pm 0.02) \times 10^{-11} $ & $(8.3 \pm 0.7) \times 10^{-12} $ & $(2.9 \pm 0.4) \times 10^{-11} $ & $(1.418 \pm 0.008) \times 10^{-11} $\\
1Jy 1406-076 & $(2.6 \pm 0.5) \times 10^{-11} $ & $(2.7 \pm 1.0) \times 10^{-13} $ & $(9 \pm 2) \times 10^{-13} $ & $(5.0 \pm 1.0) \times 10^{-13} $\\
H 1426+428 & $(3.87 \pm 0.02) \times 10^{-11} $ & $(1.92 \pm 0.11) \times 10^{-11} $ & $(1.8 \pm 0.3) \times 10^{-10} $ & $(7.37 \pm 0.04) \times 10^{-11} $\\
PKS 1510-089 & $(4.96 \pm 0.05) \times 10^{-11} $ & $(6.1 \pm 0.7) \times 10^{-12} $ & $(1.9 \pm 0.4) \times 10^{-11} $ & $(9.77 \pm 0.10) \times 10^{-12} $\\
PKS 1622-297 & $(6.0 \pm 0.3) \times 10^{-11} $ & $(1.2 \pm 0.3) \times 10^{-12} $ & $(6.4 \pm 0.7) \times 10^{-12} $ & $(2.8 \pm 0.2) \times 10^{-12} $\\
1Jy 1633+38 & $(4.97 \pm 0.12) \times 10^{-11} $ & $(1.3 \pm 0.2) \times 10^{-12} $ & $(2.22 \pm 0.09) \times 10^{-11} $ & $(3.86 \pm 0.10) \times 10^{-12} $\\
Mrk 501 & $(4.142 \pm 0.012) \times 10^{-11} $ & $(8.9 \pm 0.3) \times 10^{-11} $ & $(6.39 \pm 0.06) \times 10^{-10} $ & $(2.000 \pm 0.006) \times 10^{-10} $\\
PKS 1730-130 & $(6.4 \pm 0.3) \times 10^{-11} $ & $(1.2 \pm 0.4) \times 10^{-12} $ & $(6.7 \pm 1.0) \times 10^{-12} $ & $(2.62 \pm 0.14) \times 10^{-12} $\\
1ES 1959+650 & $(5.440 \pm 0.011) \times 10^{-11} $ & $(9.1 \pm 0.2) \times 10^{-11} $ & $(7.40 \pm 0.05) \times 10^{-10} $ & $(3.535 \pm 0.009) \times 10^{-10} $\\
PKS 2155-304 & $(4.618 \pm 0.012) \times 10^{-11} $ & $(1.74 \pm 0.11) \times 10^{-11} $ & $(6.72 \pm 0.03) \times 10^{-10} $ & $(1.077 \pm 0.003) \times 10^{-10} $\\
BL Lacertae & $(6.03 \pm 0.05) \times 10^{-11} $ & $(1.10 \pm 0.04) \times 10^{-11} $ & $(5.2 \pm 0.2) \times 10^{-11} $ & $(1.83 \pm 0.02) \times 10^{-11} $\\
3C 454.3 & $(7.34 \pm 0.02) \times 10^{-11} $ & $(6.6 \pm 1.2) \times 10^{-12} $ & $(3.60 \pm 0.05) \times 10^{-10} $ & $(7.50 \pm 0.02) \times 10^{-11} $\\
1ES 2344+514 & $(5.83 \pm 0.06) \times 10^{-11} $ & $(9.8 \pm 0.8) \times 10^{-12} $ & $(1.24 \pm 0.03) \times 10^{-10} $ & $(2.63 \pm 0.03) \times 10^{-11} $\\
\enddata
\tablecomments{Values are calculated for data through August 31st, 2012. Sources are ordered by increasing R.A..}
\end{deluxetable}

\begin{deluxetable}{ll}
\tabletypesize{\scriptsize}
\tablecaption{Light Curve Fields\label{lcFields}}
\tablewidth{0pt}
\tablecolumns{2}
\tablehead{
\colhead{Field} & 
\colhead{Description}
}
\startdata
Bin Center & The center of the time bin in units of MJD.\\
BinHW & One-half of the full time bin in units of days.\\
Rate & The background subtracted, PSF corrected and pile-up corrected source count rate in units of $counts \enskip s^{-1}$.\\
RateError & The $1\sigma$ error in the source count rate in units of $counts \enskip s^{-1}$.\\
FracExp & Fractional exposure time for the full time bin.\\
Mode & 0 if in Windowed Timing mode.\\
         & 1 if in Photon Counting mode.\\
RawRate\tablenotemark{*} & The raw source count rate in units of $counts \enskip s^{-1}$.\\
RawError\tablenotemark{*} & The $1\sigma$ error in the raw count rate in units of $counts \enskip s^{-1}$.\\
NetRate\tablenotemark{*} & The background subtracted count rate in units of $counts \enskip s^{-1}$.\\
NetError\tablenotemark{*} & The $1\sigma$ error in the net count rate in units of $counts \enskip s^{-1}$.\\
BackRate\tablenotemark{*} & The background count rate in units of $counts \enskip s^{-1}$.\\
BackError\tablenotemark{*} & The $1\sigma$ error in the background count rate in units of $counts \enskip s^{-1}$.\\
\enddata
\tablenotetext{*}{These fields are only included in the detailed light curve text files.}
\end{deluxetable}

\begin{deluxetable}{ll}
\tabletypesize{\scriptsize}
\rotate
\tablecaption{Hardness Ratio Fields\label{hrFields}}
\tablewidth{0pt}
\tablecolumns{2}
\tablehead{
\colhead{Field} & 
\colhead{Description}
}
\startdata
Bin Center & The center of the time bin in units of MJD.\\
BinHW & One-half of the full time bin in units of days.\\
FracExp & Fractional exposure time for the full time bin.\\
Ratio & Simple source hardness ratio: $R_2$/$R_1$, where $R_2$ is the rate in the 2-10 keV band and $R_1$ is the rate in the 0.3-2 keV band.\\
RatioError & $1\sigma$ error in the source hardness ratio.\\
Mode & 0 if in Windowed Timing mode\\
         & 1 if in Photon Counting mode\\
Rate1 & The background subtracted, PSF corrected and pile-up corrected source count rate in the $0.3 - 2.0$ keV band in units of $counts \enskip s^{-1}$.\\
RateError1 & The $1\sigma$ error in the source count rate in the $0.3 - 2.0$ keV band in units of $counts \enskip s^{-1}$.\\
NetRate1 & The background subtracted count rate in the $0.3 - 2.0$ keV band in units of $counts \enskip s^{-1}$.\\
NetError1 & The $1\sigma$ error in the net count rate in the $0.3 - 2.0$ keV band in units of $counts \enskip s^{-1}$.\\
Rate2 & The background subtracted, PSF corrected and pile-up corrected source count rate in the $2.0 - 10.0$ keV band in units of $counts \enskip s^{-1}$.\\
RateError2 & The $1\sigma$ error in the source count rate in the $2.0 - 10.0$ keV band in units of $counts \enskip s^{-1}$.\\
NetRate2 & The background subtracted count rate in the $2.0 - 10.0$ keV band in units of $counts \enskip s^{-1}$.\\
NetError2 & The $1\sigma$ error in the net count rate in the $2.0 - 10.0$ keV band in units of $counts \enskip s^{-1}$.\\
\enddata
\end{deluxetable}

\begin{deluxetable}{lr}
\tabletypesize{\scriptsize}
\tablecaption{Other Sources of Interest\label{publicSources}}
\tablewidth{0pt}
\tablecolumns{2}
\tablehead{
\colhead{Source Name} & 
\colhead{Total XRT Exposure Time} \\
\colhead{} & 
\colhead{(ks)}
}
\startdata
1ES 0033+595 & 22\\
TXS 0059+581 & 27\\
J0109+6134 & 41\\
4C 31.03 & 6\\
4U 0115+63 & 18\\
QSO B0133+47 & 59\\
CGRaBS J0211+1051 & 19\\
S3 0218+35 & 15\\
3C 66A & 81\\
1ES 0229+200 & 24\\
PKS 0235-618 & 18\\
4C +28.07 & 29\\
PKS 0244-470 & 7\\
PKS 0250-225 & 9\\
PKS 0301-243 & 25\\
3C 84 & 67\\
RX J0324.6+3410 & 186\\
0336-019 & 26\\
PKS 0402-362 & 45\\
PKS 0405-385 & 22\\
1ES 0414+009 & 12\\
0420-014 & 57\\
PKS 0426-380 & 24\\
NRAO 190 & 20\\
PKS 0447-439 & 50\\
PKS 0454-234 & 34\\
PKS 0458-02 & 36\\
PKS 0502+049 & 7\\
1ES 0502+675 & 40\\
VER J0521+212 & 18\\
PKS 0521-36 & 41\\
PMN J0531-4827 & 65\\
QSO B0529+075 & 31\\
PKS 0537-441 & 164\\
1A 0535+262 & 95\\
B2 0619+33 & 38\\
PKS 0646-306 & 6\\
RX J0648.7+1516 & 11\\
1ES 0647+250 & 53\\
B3 0650+453 & 16\\
RGB J0710+591 & 21\\
PKS 0722+145 & 52\\
PKS 0727-115 & 53\\
0735+178 & 41\\
87GB 073840.5+545138 & 5\\
Swift J0746.3+2548 & 107\\
PKS 0805-07 & 41\\
1ES 0806+524 & 53\\
0829+046 & 47\\
0836+710 & 64\\
PKS B0906+015 & 26\\
J0910-5041 & 22\\
QSO B0917+449 & 14\\
PMN J0948+0022 & 99\\
0954+658 & 64\\
1ES 1011+496 & 54\\
GB6 J1032+6051 & 39\\
S4 1030+61 & 39\\
1055+018 & 29\\
Swift J1112.2-8238 & 27\\
PKS 1118-056 & 9\\
PMN J1123-6417 & 17\\
PKS 1124-186 & 34\\
1127-145 & 50\\
SBS 1150+497 & 55\\
1156+295 & 56\\
1ES 1215+303 & 42\\
1ES 1218+304 & 32\\
PKS 1222+216 & 130\\
3EG J1236+0457 & 5\\
GX 304-1 & 126\\
PSR B1259-63 & 62\\
1308+326 & 40\\
GB6 B1310+4844 & 45\\
PKS 1329-049 & 22\\
B3 1343+451 & 12\\
PKS 1424+240 & 27\\
PKS 1424-418 & 36\\
1ES 1440+122 & 7\\
PKS 1440-389 & 18\\
PKS 1454-354 & 15\\
PKS 1502+106 & 88\\
QSO B1514-24 & 33\\
Cir X-1 & 142\\
B2 1520+31 & 23\\
PG 1553+113 & 111\\
1611+343 & 30\\
PKS 1622-253 & 6\\
PMN J1626-2426 & 7\\
3C 345 & 77\\
GB6 J1700+6830 & 4\\
B3 1708+433 & 5\\
PMN J1717-5155 & 20\\
QSO B1727+502 & 5\\
1ES 1741+196 & 12\\
S4 1749+70 & 44\\
OT081 & 48\\
S5 1803+784 & 55\\
PKS 1830-21 & 33\\
CGRaBS J1848+3219 & 37\\
CGRaBS J1849+6705 & 45\\
SS 433 & 24\\
PMN J1913-3630 & 1\\
IGR J19294+1816 & 54\\
XTE J1946+274 & 22\\
MG4 J200112+4352 & 62\\
PKS 2023-07 & 31\\
Cyg X-3 & 146\\
Swift J2058+0516 & 89\\
V407 Cyg & 93\\
S3 2141+17 & 69\\
PKS 2149-306 & 35\\
NRAO 676 & 21\\
3C 446 & 35\\
CTA102 & 131\\
PKS 2233-148 & 29\\
B3 2247+381 & 47\\
PMN J2250-2806 & 13\\
B2 2308+34 & 16\\
PKS 2326-502 & 32\\
PMN J2345-1555 & 36\\
\enddata
\tablecomments{Values are calculated for data from December 17th, 2004 through August 31st, 2012. Sources are ordered by increasing R.A..}
\end{deluxetable}

\clearpage

\clearpage

\clearpage

\clearpage

\clearpage

\clearpage

\clearpage

\clearpage

\clearpage

\clearpage

\clearpage

\clearpage

\clearpage

\clearpage

\clearpage

\clearpage

\clearpage

\clearpage

\clearpage

\clearpage

\clearpage

\clearpage

\clearpage

\end{document}